\documentclass[12pt]{elsarticle}
\usepackage{amssymb}
\usepackage{amsmath}
\usepackage{bm}
\biboptions{sort&compress}
\begin{document}
%\linenumbers

\begin{frontmatter}

\title{Kernel based unfolding of data obtained from detectors
  with finite resolution and limited acceptance}

%\title{Unfolding as non-parametric fit of data obtained from detectors with
%  finite resolution and limited acceptance.  I. Method description and unfolding
%  for one dimensional distributions.}

\author{N.D. Gagunashvili\corref{cor1}}
%\fnref{fn1}}
\ead{nikolai@unak.is}
\cortext[cor1]{Tel.: +354-4608505; fax: +354-4608998}
\address{University of Akureyri, Borgir, v/Nordursl\'od, IS-600 Akureyri, Iceland}
\author{M. Schmelling}
\address{Max-Planck-Institut f\"{u}r Kernphysik, P.O. Box 103980, 69029 Heidelberg,
Germany}
\begin{abstract}
  A kernel based procedure for correcting experimental data for
  distortions due to the finite resolution and limited detector acceptance
  is presented. The unfolding problem is known to be an ill-posed problem
  that can not be solved without some a priori information about solution
  such as, for example, smoothness or positivity. In the approach presented
  here the true distribution is estimated by a weighted sum of kernels,
  with the width of the kernels acting as a regularization parameter
  responsible for the smoothness of the result. Cross-validation is used
  to determine an optimal value for this parameter. A numerical example with
  a simulation study of systematical and statistical errors is presented to
  illustrate the procedure.
\end{abstract}
\end{frontmatter}

\section{Introduction}
%---------------------
In this paper the 1-dimensional unfolding problem will be addressed. Here
the probability density function (PDF) $P(x')$\/ of an experimentally measured
characteristic $x'$\/ in general differs from the true physical PDF $p(x)$\/
because of the limited acceptance (probability) $A(x)$\/ to register
an event with true characteristic $x$\/ and finite resolution in the response
function $R(x'|x)$, the probability to observe $x'$\/ for a given true
value $x$. Formally the relation between $P(x')$\/ and $p(x)$\/ is
given by
\begin{equation}
    P(x') \propto \int_{\Omega} p(x)A(x)R(x'|x) \,dx \;.
\label{p1_main}
\end{equation}
The integration in (\ref{p1_main}) is carried out over the domain
$\Omega$ of the variable $x$. In practical applications the experimental
distribution is usually discretized by using a histogram representation,
obtained by integrating $P(x')$\/ over $n$\/ finite size bins
\begin{equation}
    P_{j} = \int_{c_{j-1}}^{c_j} P(x') dx' \quad j=1,\ldots,n
\end{equation}
with $c_{j-1}, c_j$\/ the bounds of bin $j$.

If a parametric (theoretical) model $p_T(x,a_{1},a_{2},\ldots, a_{l})$\/ for the true
PDF is known, then the unfolding can be done by determining the parameters in
a least squares fit to the binned data \cite{fitgagunash} or a maximum
likelihood fit to the unbinned data. In both cases the a priori information
which is needed to correct for the distortions by the experimental setup is
the fit model, which allows to describe the true distribution by a
finite number of parameter values.

Model independent unfolding to identify a physical distribution,
as considered in \cite{zhigunov2,blobel,correcting,gagunashvili,schmelling,
hocker,zech,agost,blobel2,gagunashvili_phystat,albert}, is an underspecified
problem and every approach to solving it requires a priori information
about the solution. Different methods differ, directly or indirectly, in the use
of this a priori information.

The remainder of the paper is organized as follows. In section \ref{sec:method}
a new method for solving the unfolding problem will be presented. Properties of
the algorithm are discussed in section \ref{sec:discuss} and illustrated in
section \ref{sec:example} by applying it to a numerical example proposed in
\cite{blobel} and also used in Refs.\,\cite{schmelling,hocker}. Conclusions
are given in section \ref{sec:conclusions}.

\section{Description of the unfolding method}
%---------------------------------------------
\label{sec:method}
To solve the unfolding problem (\ref{p1_main}) the following ansatz
for $p(x)$\/ will be used
\begin{equation}
  p(x)= w_0+ \sum_{i=1}^{s} w_i \, K(x,x_i,\lambda), \label{repres}
\end{equation}
where the true distribution is written as an offset $w_0$\/ plus a weighted
sum of $s$\/ kernel functions (PDFs) $K(x,x_i,\lambda),\,i=1,\ldots s$, with
non-negative weights $w_i$, central locations $x_i$\/ and a scale parameter
$\lambda$\/ which determines the width of the kernel. Kernels are widely used
for the estimation of a PDF \cite{silverman} as well as in non-parametric
regression analysis \cite{hardle}. Note that Eq.(\ref{repres}) uses
only kernels of one type with a common scale parameter. The only difference
between different kernels is the location of the center. In this paper we will
only consider this simplified case. In principle the approach could be
generalized to vary also functional form and scale parameter.

Using Eq.(\ref{repres}) to parametrize the solution $p(x)$\/ reduces the
unfolding problem of finding a solution from the infinitely many dimensional
space of all functions to finding a solution in a finite dimensional space.
This way a discretization is performed which, in contrast to e.g. a
discretization by a histogram, has the advantage to introduce negligible
quantization errors for sufficiently smooth distributions.

The following discussion will focus on symmetric kernels, although, depending on
the kind of problem one attempts to solve, also asymmetric kernels may be
appropriate. The a priori information of $p(x)$\/ being proportional to a PDF
is incorporated by accepting only positive weights. The scale parameter of
the kernel functions acts as a regularization parameter which allows to
adjust the smoothness of the result. Weights, locations and the number of
kernel functions needed to estimate $p(x)$\/ will be determined by
the unfolding procedure described below.

Below examples of smooth symmetric kernels $K(x_i-x)=K(x_i+x)$\/ are presented.
All kernels are PDFs which are normalized to unity when integrating over $x$.
For convenience, in all cases the  variable  $u=(x-x_i)/\lambda$\/ is used.
With the indicator function
\begin{equation*}
    I_{\{\cdots\}} = \left\{ \begin{array}{l}
                            1 \qquad \mbox{if $u$\/ satisfies the condition in
                                           the brackets}\\
                            0 \qquad \mbox{otherwise}
                           \end{array}\right.
\end{equation*}
a class of polynomial kernels is defined by
\begin{equation}
    K(u,\lambda) = \frac{N(a,b)}{\lambda} (1 - |u|^a)^b \,I_{\{|u|\leq 1\}} \;.
\end{equation}
Often used are the following special cases:
\begin{center}
\begin{tabular}{lccc}
  kernel       & $a$ & $b$ & $N(a,b)$ \\
\hline
  Epanechnikov &  2  &  1  &  4/3     \\
  Biweight     &  2  &  2  &  15/16   \\
  Triweight    &  2  &  3  &  35/32   \\
  Tricube      &  3  &  3  &  70/81   \\
\end{tabular}
\end{center}
Commonly employed non-polynomial kernels are:
\begin{eqnarray}
  \mbox{Cosine:}
&& K(u,\lambda) = \frac{\pi}{4 \lambda} \cos(\frac{\pi u}{2})\,I_{\{|u|\leq 1 \}}\\
  \mbox{Cauchy:}
&& K(u,\lambda) = \frac{1}{\lambda \pi}(\frac{1}{1 +u^2})\\
  \mbox{Gaussian:}
&& K(u,\lambda) = \frac{1}{\lambda \sqrt{2\pi}}e^{-\frac{u^2}{2}}
\end{eqnarray}
Also frequently used is the piecewise defined cubic B-spline
\begin{equation}
 K(u,\lambda)
 = \frac{1}{3 \lambda}
   \left\{ \begin{array}{l}
            (2u+2)^3\,I_{\{-1 \leq u <-0.5\}}                \\[1mm]
            (1+3(2u+1)(1-2u(2u+1)))\,I_{\{ -0.5 \leq u < 0\}}\\[1mm]
            (1-3(2u-1)(1-2u(2u-1)))\,I_{\{ 0 \leq u < 0.5\}} \\[1mm]
            (2-2u)^3\,I_{\{ 0.5 \leq u < 1\}}
           \end{array}\right.\;.
\end{equation}

Re-writing Eq.(\ref{repres}) in the form
\begin{equation}
   p(x) = \sum_{i=0}^s w_i \, K_i(x)
   \quad\mbox{with}\quad
   K_i(x) = \left\{ \begin{array}{l}
                   1  \quad\qquad\qquad \mbox{for $i=0$} \\
                   K(x,x_i,\lambda) \quad \mbox{for $i>0$}
                   \end{array}\right.
\end{equation}
and substituting this into the basic equation (\ref{p1_main}) yields
\begin{equation}
P(x')=  \sum_{i=0}^{s} w_i \; \int_{\Omega} K_i(x) A(x)R(x'|x) \,dx
\end{equation}

Taking statistical fluctuations into account, the relation between
the weights $w_i$\/ and the histogram of the observed distribution becomes
a linear equation
\begin{equation}
 \bm{P} =  \bm{\mathrm{Q}{w}}+  \bm{\epsilon}\, , \label{basicp}
\end{equation}
where $\bm{P}$ is the $n$-component column vector of the experimentally
measured histogram, $\bm{w} = (w_0,w_1,...,w_s)'$ \/is ($s+1$)-component
vector of  weights and $\bm{\mathrm{Q}}$\/ is an $n \times (s+1)$\/ matrix
with elements
\begin{equation}
   Q_{ji} = \int_{c_{j-1}}^{c_j} K_i(x)\,A(x)\,R(x'|x)
   \quad j=1,\ldots n \quad i=0,\ldots s \;.
\end{equation}

The vector $\bm{\epsilon}$\/ is an $n$-component vector of random residuals with
expectation value $E[\bm{\epsilon}]=\bm{0}$\/ and covariance matrix $\bm{C}$\/
with diagonal elements
$\mathrm{Var}[\bm{\epsilon}]=\mathrm{diag}(\sigma_1^2,\sigma_2^2,\cdots,\sigma_n^2)$,
where $\sigma_i$\/ is the statistical error of the measured distribution for the
$i$th bin. Each column of matrix $\bm{\mathrm{Q}}$\/ is the response of the
system to the true distribution represented by the respective kernel.
Numerically the calculation of the column vectors can be done by weighting the
events of a Monte Carlo sample such that they follow the distribution the
corresponding kernel, see Ref. \cite{fitgagunash}, and taking the histogram of
the observed distribution obtained with the weighted entries.

For a given set of kernels the weights $\bm{w}$\/ in Eq.(\ref{basicp}) can be
determined by a linear least squares fit. In order to have an as flexible as
possible model, the candidate kernels in principle could have a continuous
range of central positions. In practical applications it will usually
be sufficient to consider a discrete set with a spacing significantly smaller
than the bandwidth $\lambda$. The goal then is to find a subset of kernels
for the final fit which provides a good description of the data and where
all weights are positive and significantly different from zero. This at the
same time stabilizes the solution and guarantees positiveness.

To find such an optimal subset, a forward stepwise algorithm
\cite{seber} is used. It requires a criterion for the quality of the fit which
will be taken the test statistic $X^2_l$,
\begin{equation}
   X^2_l = (\bm{P} -\bm{Q} \hat{\bm{w}})^T \bm{C}^{-1} (\bm{P} - \bm{Q}\hat{\bm{w}})
\end{equation}
where the index $l$\/ denotes the number of weights in the fit
and $\hat{\bm{w}}$\/ is determined such that it minimizes $X^2_l$.
The solution $\hat{w}$\/ and its covariance matrix $\bm{C}_w$\/ are
given by the well known expressions
\begin{equation}
\label{eq:lsq}
  \hat{\bm{w}} = (\bm{Q^T C^{-1} Q})^{-1}\,(\bm{Q^T C^{-1}})\, \bm{P}
  \quad\mbox{and}\quad
  \bm{C}_w = (\bm{Q^T C^{-1} Q})^{-1} \;.
\end{equation}
If the underlying distribution of the measured histogram $\bm{P}$\/
can be described by a linear combination of the columns of $\bm{Q}$,
then the  $X^2_{l}$ statistics follows a $\chi^2$-distribution with
$n-l$\/ degrees of freedom.

Now assume a total of $s$\/ candidate kernel function
$K_i(x), i=1,\ldots,s$\/ with centers evenly spaced along the
possible values $x$\/ of the true distribution. In a first step
the weight $\hat{w}_0$\/ is determined by fitting only the
constant function $K_0$\/ to the data. Then an iterative procedure
starts with alternating ``Forward'' and ``Backward'' steps described
below.

Given a fit model consisting of $l$\/ kernels, in the next Forward step
each of the other $s-l$\/ kernels is tried for inclusion into the model.
From all combinations that one is selected where all weights are positive
and which gives the largest reduction in $X^2_l$. If no such fit is found
then the procedure stops. Otherwise the new kernel is included into the
model if
\begin{equation}
  \frac{X_l^2-X_{l+1}^2}{X_{l+1}^2}(n-l-1)>F_{in} \;, \label{eneq}
\end{equation}
i.e. if the reduction in $\chi^2$\/ is sufficiently large. Also in case
the best fit does not satisfy Eq.(\ref{eneq}) the procedure stops.
After accepting a new kernel into the model a Backward step is performed.
Here in turn each of the previously included kernels is removed from
the model and the test-model fitted to the data. From all fits which
have only positive weights the one with the smallest increase in
$X^2_l$\/ is taken. If the increase is below a certain threshold
\begin{equation}
  \frac{X_{l-1}^2-X_{l}^2}{X_{l}^2}(n-l)< F_{out} \label{eneqout}
\end{equation}
the respective kernel is removed from the model, and the Backward step
is iterated with the reduced model. If no kernel is removed then again a
Forward Step is tried. The procedure stops if neither a Forward, nor a
Backward Step can be done.

For the stepwise method defined above, appropriate thresholds $F_{in}$\/ and
$F_{out}$\/ must be chosen. Usually one uses $F_{in}=F_{out}=F_0$. There is no
common opinion about the best value for this constant. Reference\,\cite{efr}
for example used $F_0=2.5$, the authors of Ref.\,\cite {drep} used $F_0=3.29$\/
for the same sample of data. To allow the inclusion of as many kernels as possible
into the model, very small values $F_0$\/ can be used.

When the method stops an estimate $\hat{p}(x)$\/ has been found, defined by
the locations $x_i, i=1\ldots,k$\/ of a set of kernel functions which are
summed with weights $w_i,i=0,\ldots,k$\/ to yield
\begin{equation}
  \hat p(x)= \sum_{i=0}^{k} \hat{w}_i K_i(x) \label{px}  \;.
\end{equation}
The error band around $\hat{p}(x)$\/ is given by
$\sqrt{\mathrm{var}[\hat{p}(x)]}$, obtained by setting $x=y$\/ in
the expression for the covariance between any two points $x$\/ and $y$\/
\begin{equation}
   \mathrm{cov}[\hat{p}(x),\hat{p}(y)]
 = \sum_{i,j=0}^k K_i(x)\, K_j(y)\, (\bm{C}_w)_{ij}\;.
\end{equation}
A histogram representation for the unfolded distribution $\hat{p}(x)$\/
with $m$\/ bins integrating over the $x$-intervals $[b_{i-1},b_i],\,i=1,\ldots,m$\/
is obtained by
\begin{equation}
  \hat{\bm{p}}=\bm{\mathrm K} \, \bm{\hat w}, \label{unfbin}
\end{equation}
where $\bm{\mathrm K}$\/ is an $m\times (k+1)$\/ matrix with elements
\begin{equation}
   \mathrm{K}_{ij} = \int_{b_{i-1}}^{b_i} K_j(x)\, dx \;.
\end{equation}
The covariance matrix of $\hat{\bm{p}}$\/ is given by
\begin{equation}
   \bm{C}_p = \bm{\mathrm{K}}^T \, \bm{C}_w \, \bm{\mathrm{K}} \;.
\end{equation}
Note that this matrix is singular when the number of weights is
smaller than the number of bins in the histogram of the unfolded
distribution.

\section{Discussion}
\label{sec:discuss}
The unfolding algorithm described above defines a generic
approach to represent measured information about a true physical distribution
in a compact way. The fact that the model is specified with proper
statistical errors allows a quantitative comparison between an
independent theoretical model and the unfolding result when working
on the subspace spanned by the model $\hat{p}(x)$. To test the hypothesis
that the underlying distribution of the unfolding result has the shape
$p_T(x)$, one can use the histogram representation of $\bm{p}_T$\/ with
the same binning as for $\hat{\bm{p}}$. In case of a non-singular
covariance matrix $\bm{C}_p$\/ a $\chi^2$-test can be applied directly
on the binned distributions. If the number of bins for the unfolded
distribution is larger than the number of weights, the comparison can
still be done in the space spanned by the weights. In this case the
weight vector $\bm{w}_T$\/ for expanding $\bm{p}_T$\/ into the kernels
$\bm{\mathrm{K}}$\/ is given by
\begin{equation}
    \bm{w}_T = (\bm{\mathrm{K}^T\mathrm{K}})^{-1}\,\bm{\mathrm{K}}\,\bm{p}_T
\end{equation}
which, in analogy to Eq.(\ref{eq:lsq}) is simply the unweighted fit
of the kernel functions used to describe the model to the theoretical
prediction $\bm{p}_T$. If $p_T(x)$\/ is indeed the underlying distribution
of the unfolding result, then the test statistic
\begin{equation}
  X^2 = (\hat{\bm{w}} - \bm{w}_T)^T\, \bm{C}_w^{-1} \,(\hat{\bm{w}} - \bm{w}_T)
\end{equation}
has a $\chi^2$-distribution with $k+1$\/ degrees of freedom, the rank of the
matrix $\bm{C}_w$. It has to be emphasized that the above test constitutes
only a necessary condition for a theoretical prediction to describe the data.
It is not a sufficient one, as examples can be constructed where additional
kernels would be needed to properly model the prediction, which may be
known to be absent in the data and thus are ignored in the test.
In practical applications one therefore also should make sure that
$\bm{\mathrm{K}}\,\bm{w}_T$\/ provides a good model for $p_T(x)$.

In principle any smooth kernels can be used and in
practice results do not vary significantly when switching between the
functions discussed before. The choice of the optimal type of kernel
function and the value of the scale parameter $\lambda$\/ for a given
problem is driven by the quality of the fit. Common tools to asses the
fit quality in regression analysis \cite{seber} are:
\begin{enumerate}
  \item $p$-value of fit
  \item analysis of the normalized residuals of the data
  \begin{enumerate}
      \item as a function of the estimated value $\hat{\bm P}$
      \item as a function of the observed value  $x'$
  \end{enumerate}
  \item Q-Q plot: quantile of normalized residuals versus the theoretical
       quantile expected from a standard normal $\mathcal{N}(0,1)$\/ distribution
\end{enumerate}

The positions of the kernels considered in the algoritm should cover the
entire allowed range of $x$\/ with a spacing significantly smaller than
the width given by the scale parameter $\lambda$. In order to avoid loss of
information due to binning the number of bins for the measured histogram
$\bm{P}$\/ should be as large as possible although, in order to
have meaningful error estimates for the least squares fits that determine
$\bm{\hat{w}}$, the number of entries in a single bin should not be
less than $\sim 25$.

An issue left open in the definition of the unfolding algorithm is
the determination of the scale parameter $\lambda$\/ of the kernel
functions. Evidently, larger values will in general result in a more
smooth estimate for the true distribution but may lead to bad fits
of the observed distribution when narrow features cannot be
accommodated. Too small values of $\lambda$, on
the other hand, will favor overfitting of statistical fluctuations in
the data. In general one will therefore try a range of values
for $\lambda$\/ and, in order to find some optimal balance between
smoothness of the result and overfitting of the data, select a parameter
in the region just below the largest value which provides a satisfactory
fit to the data. In the literature \cite{cook, nancy, golub,allen} the
use of cross-validation or bootstrap methods is suggested to find the
optimal solution. Here we will use a simple leave-one-out cross-validation
approach \cite{allen} to determine the best value for $\lambda$.

Finally it should be noted  that the unfolding method described above does
not take into account uncertainties in the matrix $\bm{Q}$\/ which relates
the weight vector $\bm{\hat{w}}$\/ to the measurements $\bm{P}$. Therefore,
when $\bm{Q}$\/ is determined by means of a Monte Carlo simulation the
Monte Carlo sample should be significantly larger than the data sample.

%An extension of the method which is applicable also in cases where the
%Monte Carlo statistics is of the same order or less than the data statistics
%will be presented in a forthcoming publication.

\section{A numerical example }
\label{sec:example}
\begin{figure}[tb]
\centering
\includegraphics[width=0.475\textwidth]{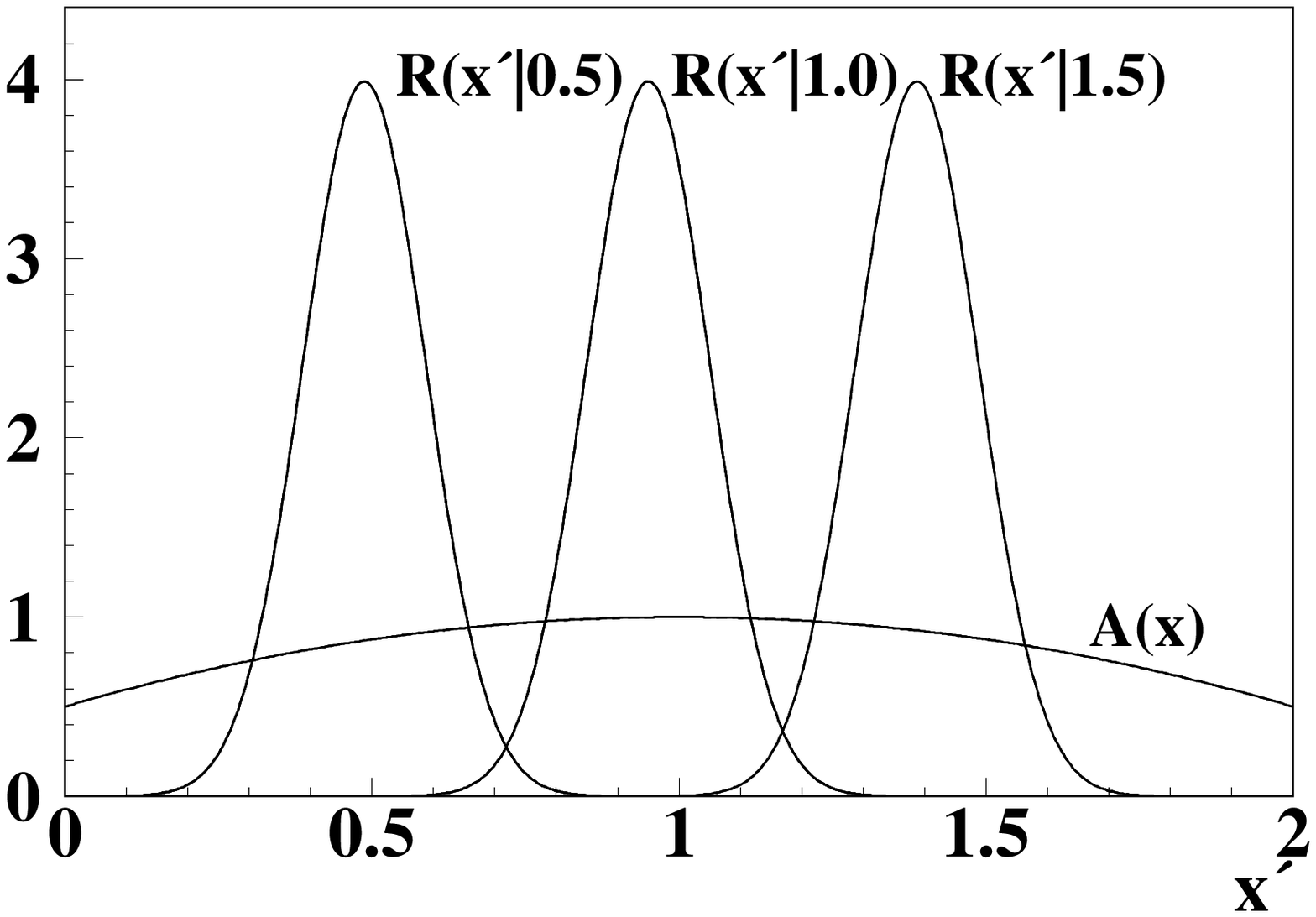}\hfill
\includegraphics[width=0.475\textwidth,height=4.7cm]{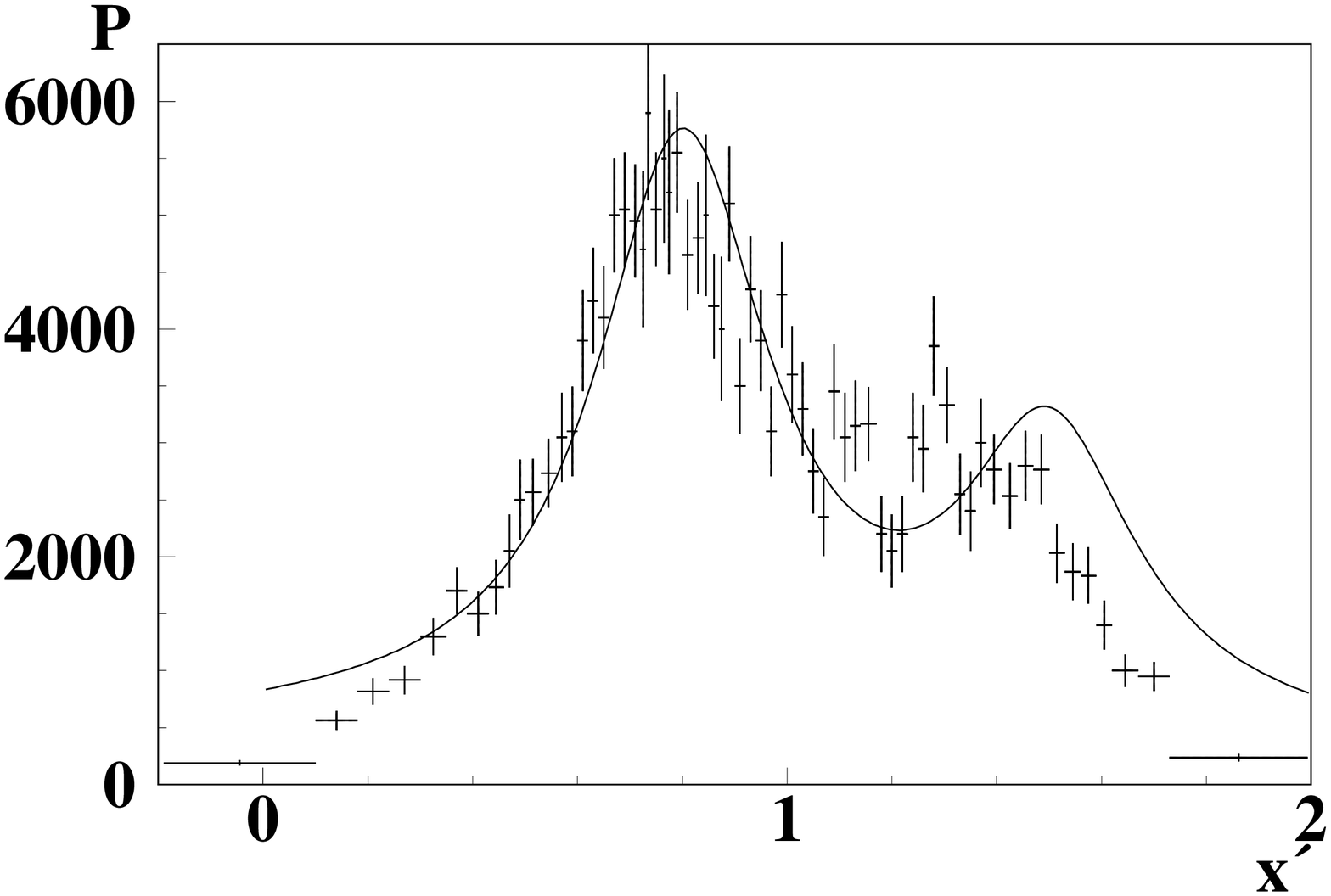}
\caption{The acceptance function $A(x)$\/ and resolution function $R(x'|x)$\/
         for $x=0.5, 1.0$ and $1.5$ (left) and histogram of the measured
         distribution $\bm{P}$\/ based on a sample of 5000 events generated
         for the true distribution (right). The true distribution $p(x)$\/
         is shown by the curve.}
\label{fig:numex1}
\end{figure}

The method described above is now illustrated with an example proposed
by Blobel \cite {blobel} and for illustration also used elsewhere
\cite{schmelling,hocker}. The true distribution, defined on the range
$x \in [0,2]$\/ is described by a sum of three Breit-Wigner functions
\begin{equation}
p(x) \propto \frac{4}{(x-0.4)^2+4}
       +     \frac{0.4}{(x-0.8)^2+0.04}
       +     \frac{0.2}{(x-1.5)^2+0.04}
\label{testform}
\end{equation}
from which the experimentally measured distribution is obtained by
\begin{equation}
P(x') \propto \int_0^{2} p(x)A(x)R(x'|x)dx ,
\end{equation}
with an acceptance function $A(x)$
\begin{equation}
A(x)=1-\frac{(x-1)^2}{2}
\end{equation}
and a resolution function describing a biased measurement with gaussian smearing
\begin{equation}
R(x'|x)=\frac{1}{\sqrt{2\pi}\sigma}\exp\left(-\frac{(x'-x+0.05x^2)^2}{2\sigma^2}\right)
\quad\mbox{with}\quad
\sigma=0.1 \;.
\end{equation}
The acceptance and resolution functions are shown in Fig.\,\ref{fig:numex1}.
Also shown is an example for the measured distribution obtained by
simulating a sample of $N=5000$\/ events.

For the determination of the matrix $\bm{Q}$\/ a sample of 500\,000 Monte
Carlo events was simulated. The true distribution was taken uniform and the
kernel responses were calculated by weighting the Monte Carlo events with
weights proportional value of kernel function \cite{fitgagunash}. A set
of 100 gaussian kernels was used with positions uniformly distributed
over the interval $[0,2]$. For the nominal analysis a scale parameter
$\lambda=0.175$\/ and a threshold value $F_0=10^{-4}$\/ in the
stepwise algorithm was chosen.

\begin{figure}[t!]
\centering
\includegraphics[width=0.6\textwidth,height=5.5cm]{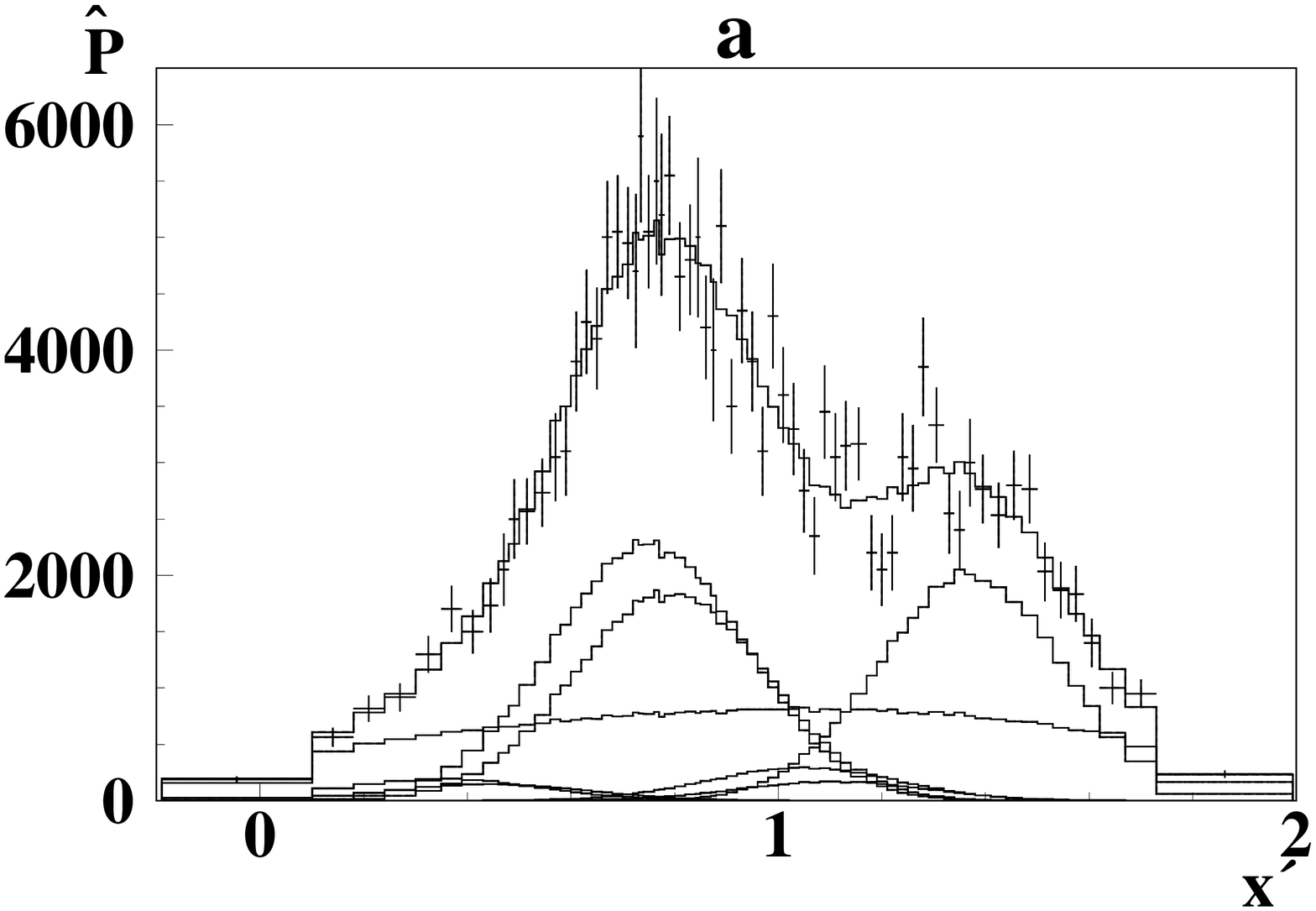}\hfill
\includegraphics[width=0.33\textwidth,height=5.6cm]{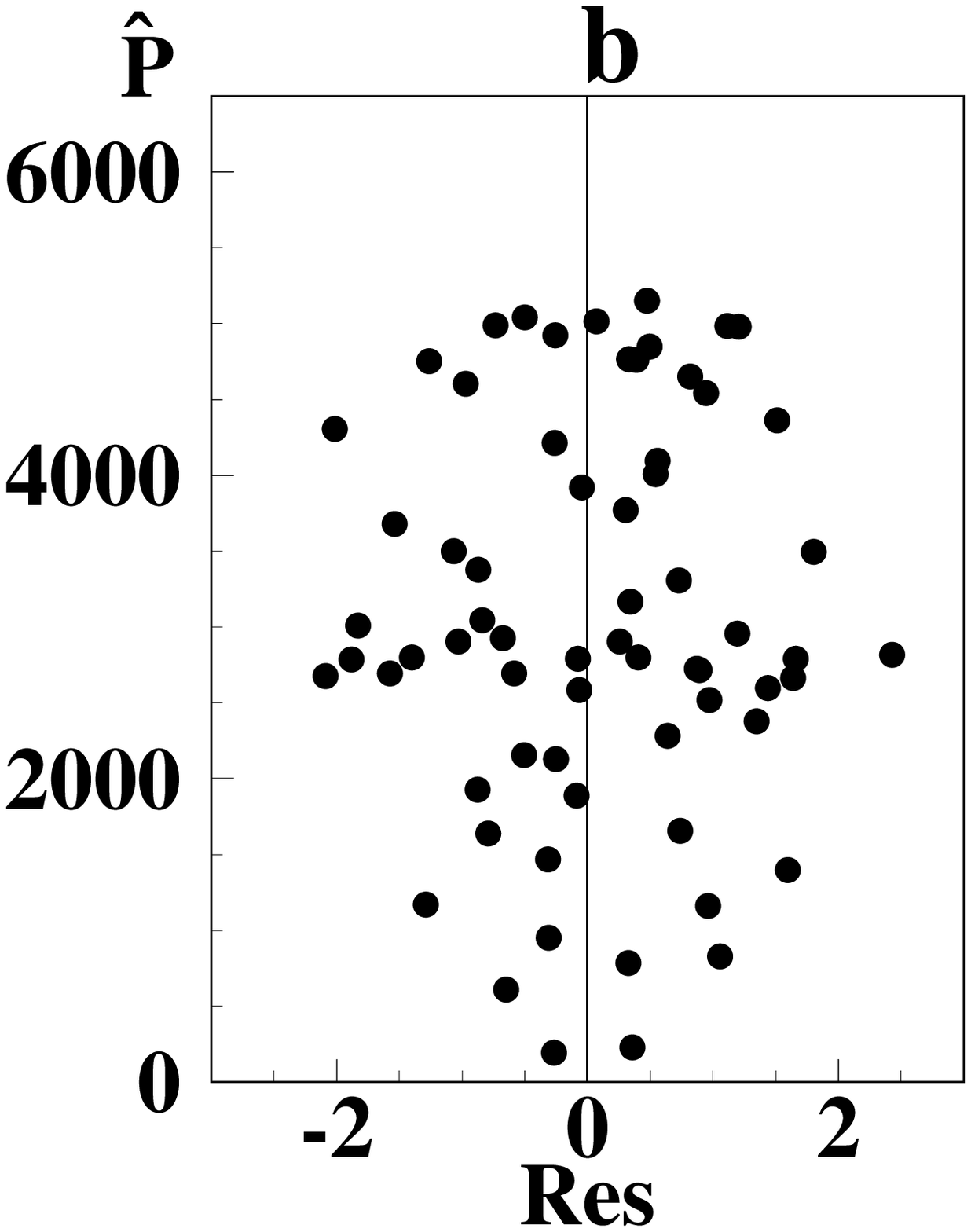}\\[5mm]
\includegraphics[width=0.6\textwidth,height=4.0cm]{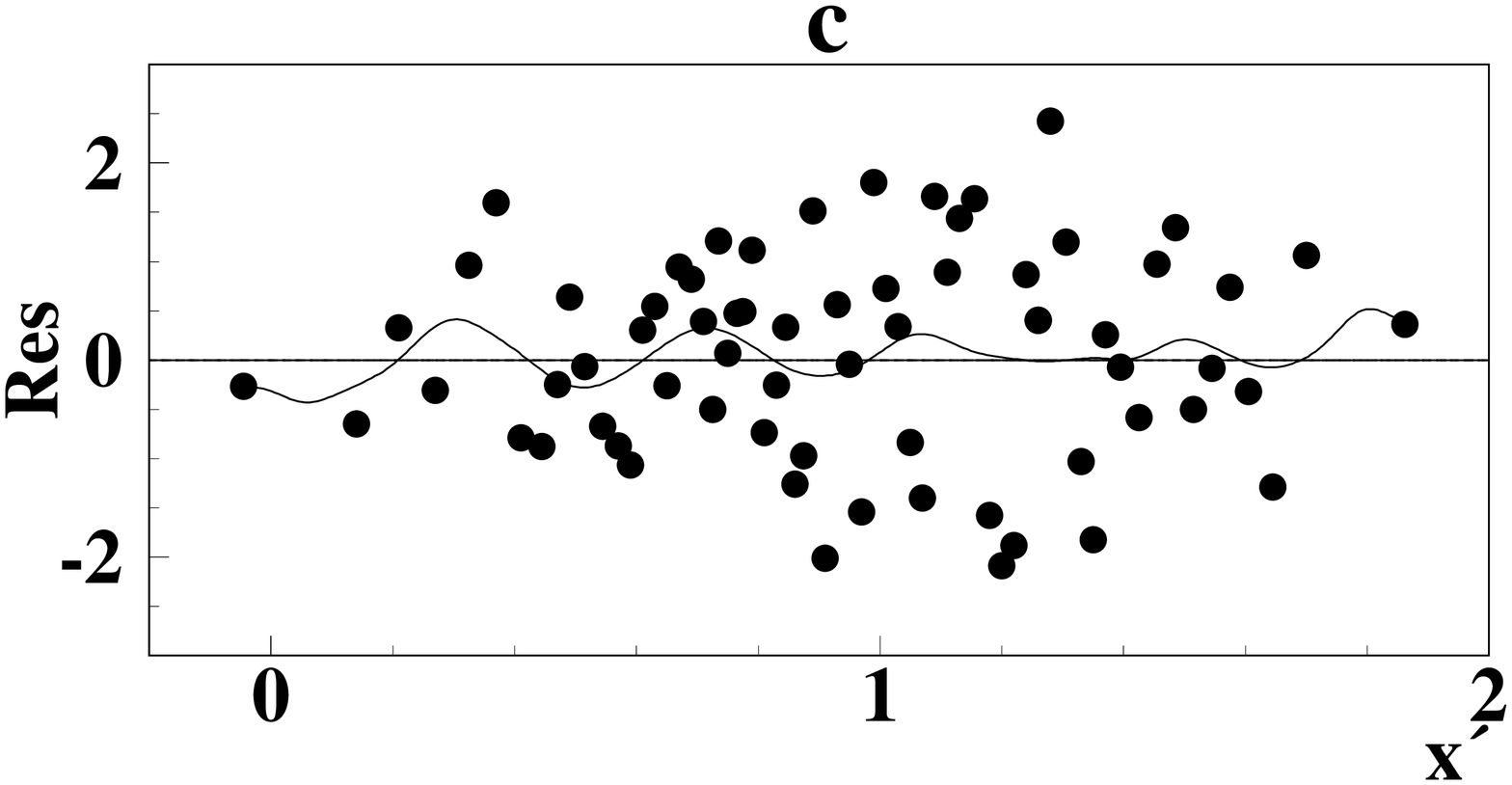}\hfill
\raisebox{-2mm}{\includegraphics[width=0.32\textwidth]{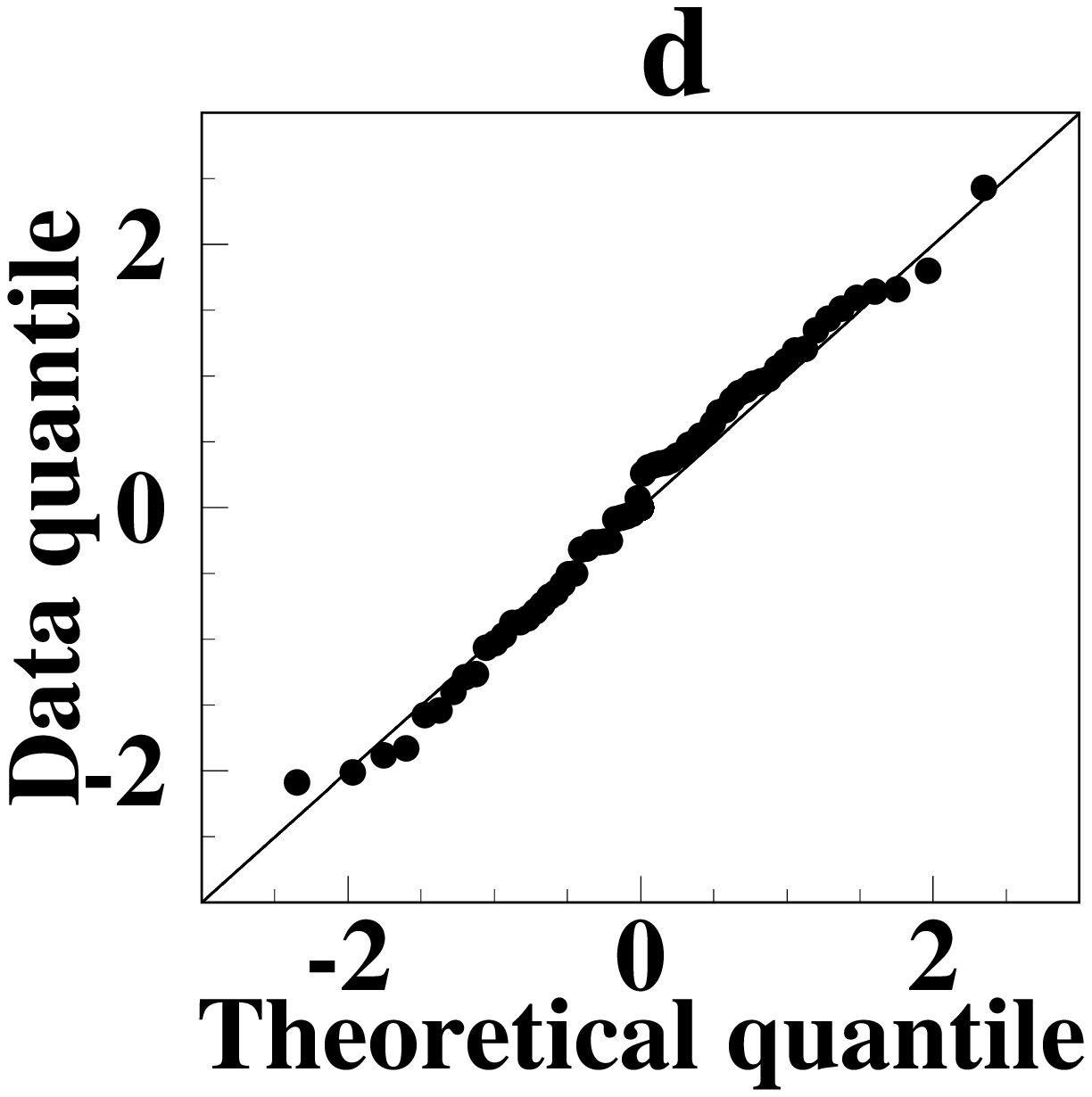}}
\caption{Illustration of the quality of the unfolding result.
        (a) folded kernels of the estimate of the true distribution
        compared to the measured distribution; (b) normalized residuals
        of the fit as a function of $\hat{\bm{P}}$; (c) normalized
        residuals as a function of $x'$; (d) quantile-quantile-plot
        for the normalized residuals.}
\label{fig:numex2}
\end{figure}

The estimate for the true distribution obtained by the unfolding method
described above is represented by a constant plus a weighted sum of seven
kernels. The positions and weights of the kernels determined by the stepwise
algorithm together with the errors and correlation matrix of the weights is
listed in Tab.\,\ref{tab:result1}. The quality of the unfolding result is
illustrated by Fig.\,\ref{fig:numex2}. It shows the superposition of the folded
kernels approximates the measured distribution together with the analysis
of the residual and the quantile-quantile plot. No structure in either of
the control plots is observed. The $p$-value from the test for the comparison
of the histogram of the measured distribution $\bm{P}$\/ and the fitted
histogram $\hat{\bm{P}}$, Fig.\,\ref{fig:numex1}(a), is $p=0.23$.

\begin{table}[tb]
\centering
\caption{Positions of kernels $x_i$, weights $\hat{w}_i$, errors of weights
         $\delta^w_i$ and correlation matrix for the weights determined by
         the unfolding algorithm.}
\label{tab:result1}
\small
\begin{tabular}{c|c|r|r|rrrrrrr}
 $i$& $x_i$ & $\hat w_i$ & $ \delta^w_i $ & 0 & 1 & 2 & 3 & 4 & 5 & 6 \\
\hline
  0&  --- &  1456.3&  268.4&      &      &      &      &      &      &      \\
  1&  0.33&   122.8&  282.6& -0.70&      &      &      &      &      &      \\
  2&  0.77&  1111.4& 1691.1& -0.51&  0.81&      &      &      &      &      \\
  3&  1.18&    79.1&  919.4&  0.54& -0.65& -0.86&      &      &      &      \\
  4&  1.11&   137.3& 1139.7& -0.55&  0.68&  0.90& -0.99&      &      &      \\
  5&  0.82&   891.3& 1816.6&  0.49& -0.78& -0.99&  0.88& -0.92&      &      \\
  6&  0.43&    85.1&  350.5&  0.53& -0.96& -0.88&  0.66& -0.70&  0.85&      \\
  7&  1.50&  1029.1&  164.4& -0.82&  0.68&  0.66& -0.82&  0.80& -0.66& -0.58
\end{tabular}
\end{table}

\begin{figure}[p]
\begin{center}$
\begin{array}{cc}
\vspace*{-1.92cm}\includegraphics[width=2.99in]{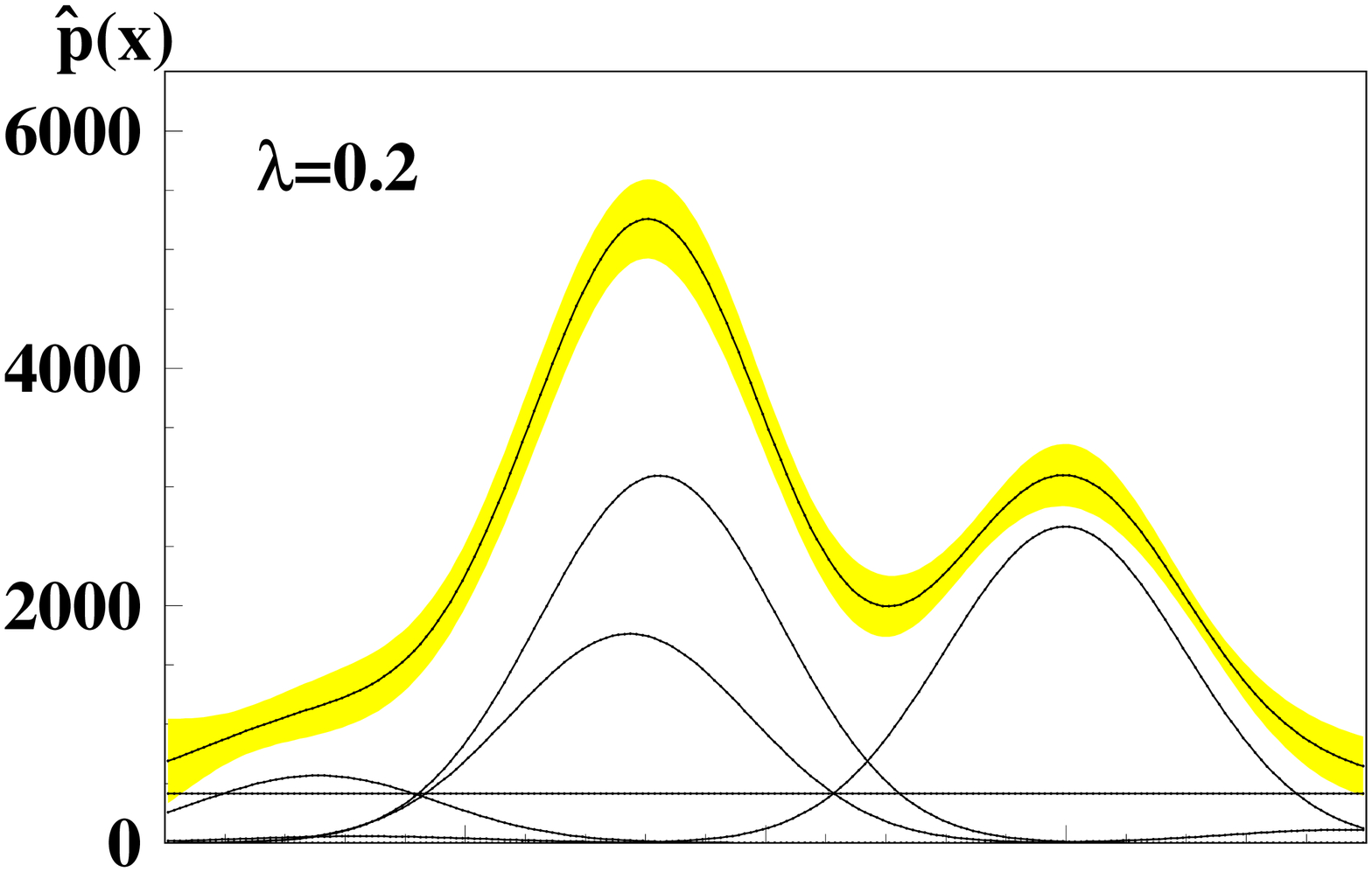} &
\hspace*{-1.82cm}\includegraphics[width=2.99in]{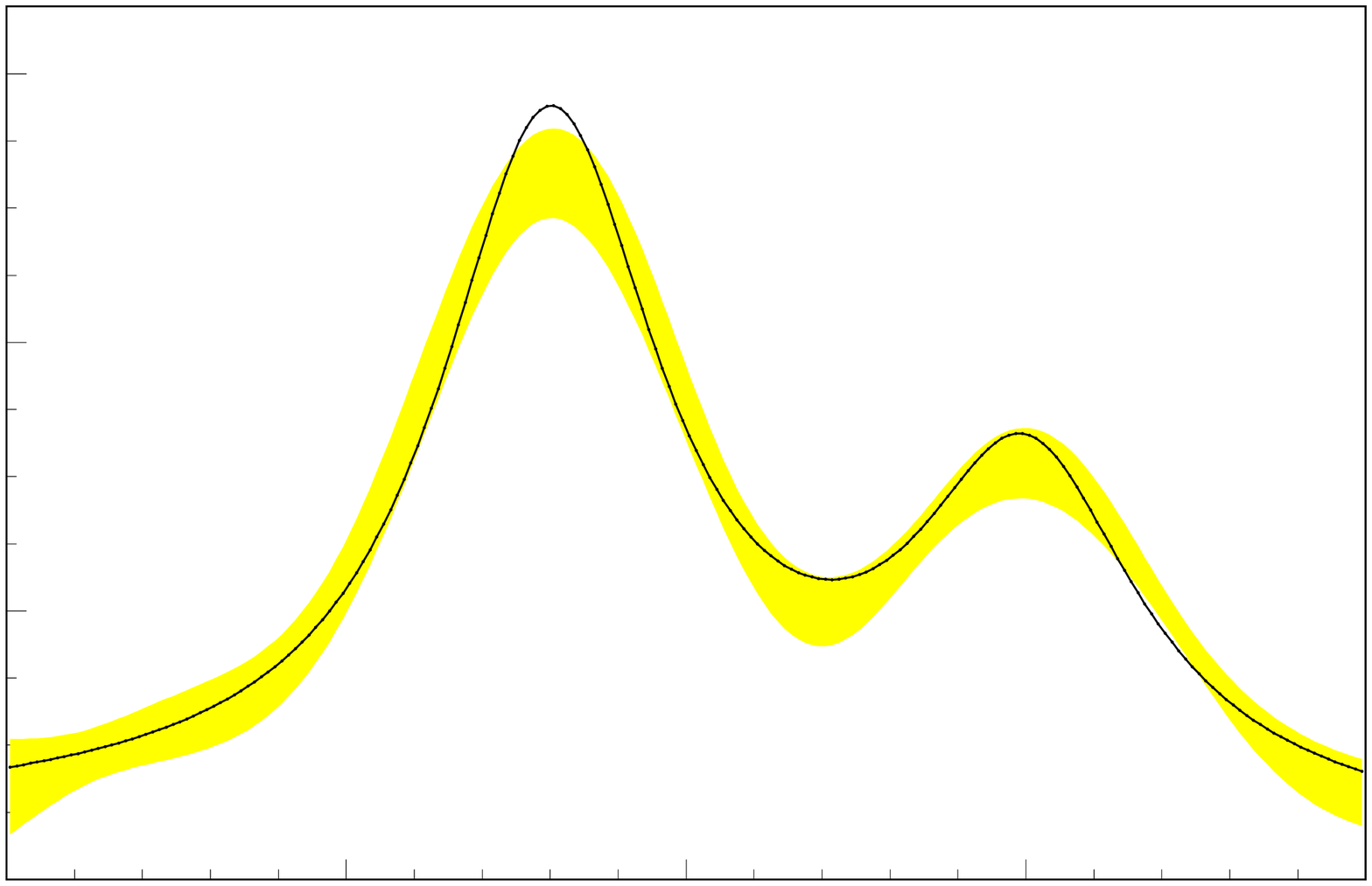}\\
\vspace*{-1.92cm}\includegraphics[width=2.99in]{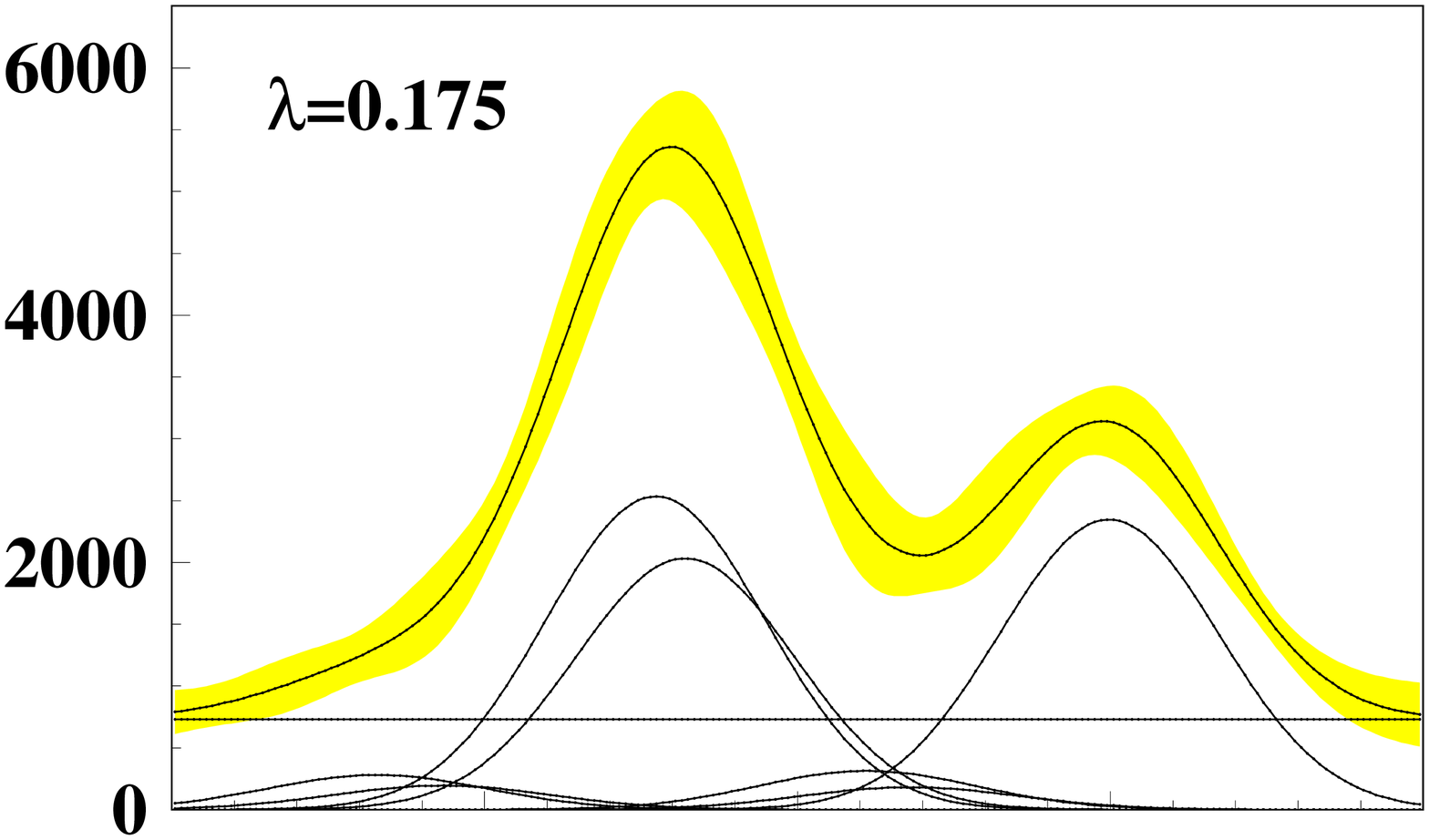} &
\hspace*{-1.82cm}\includegraphics[width=2.99in]{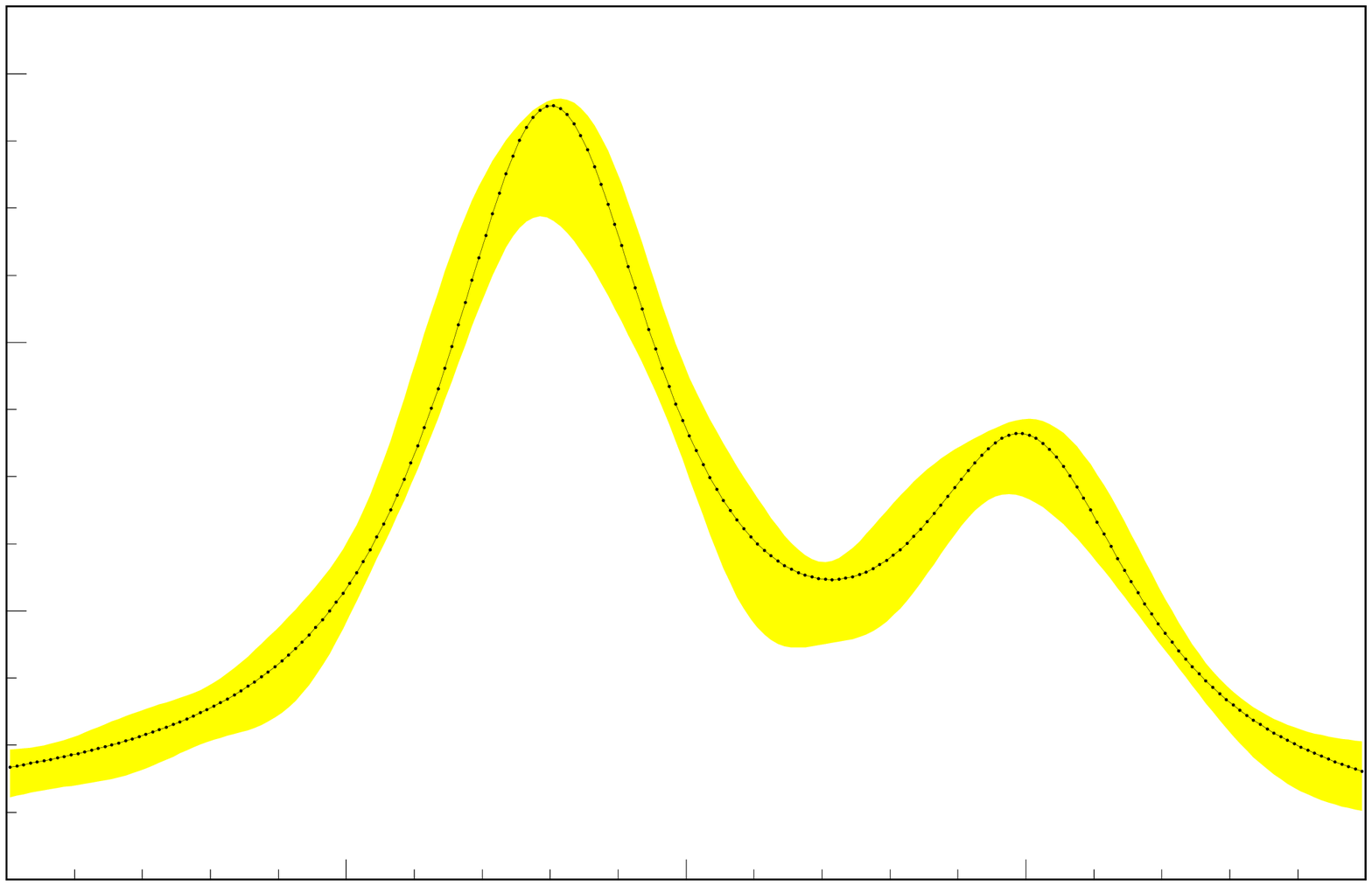}\\
\vspace*{-1.92cm}\includegraphics[width=2.99in]{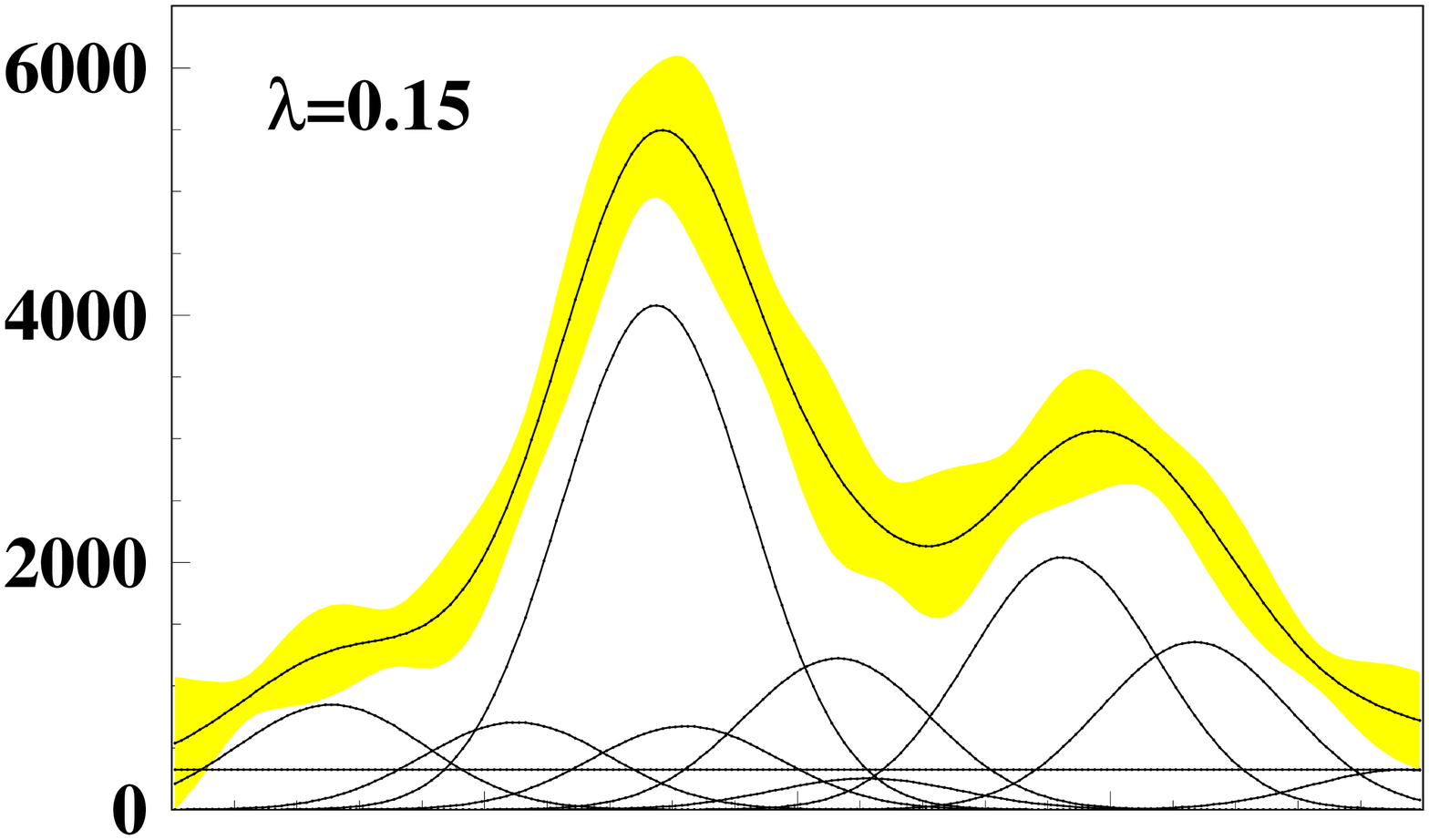} &
\hspace*{-1.82cm}\includegraphics[width=2.99in]{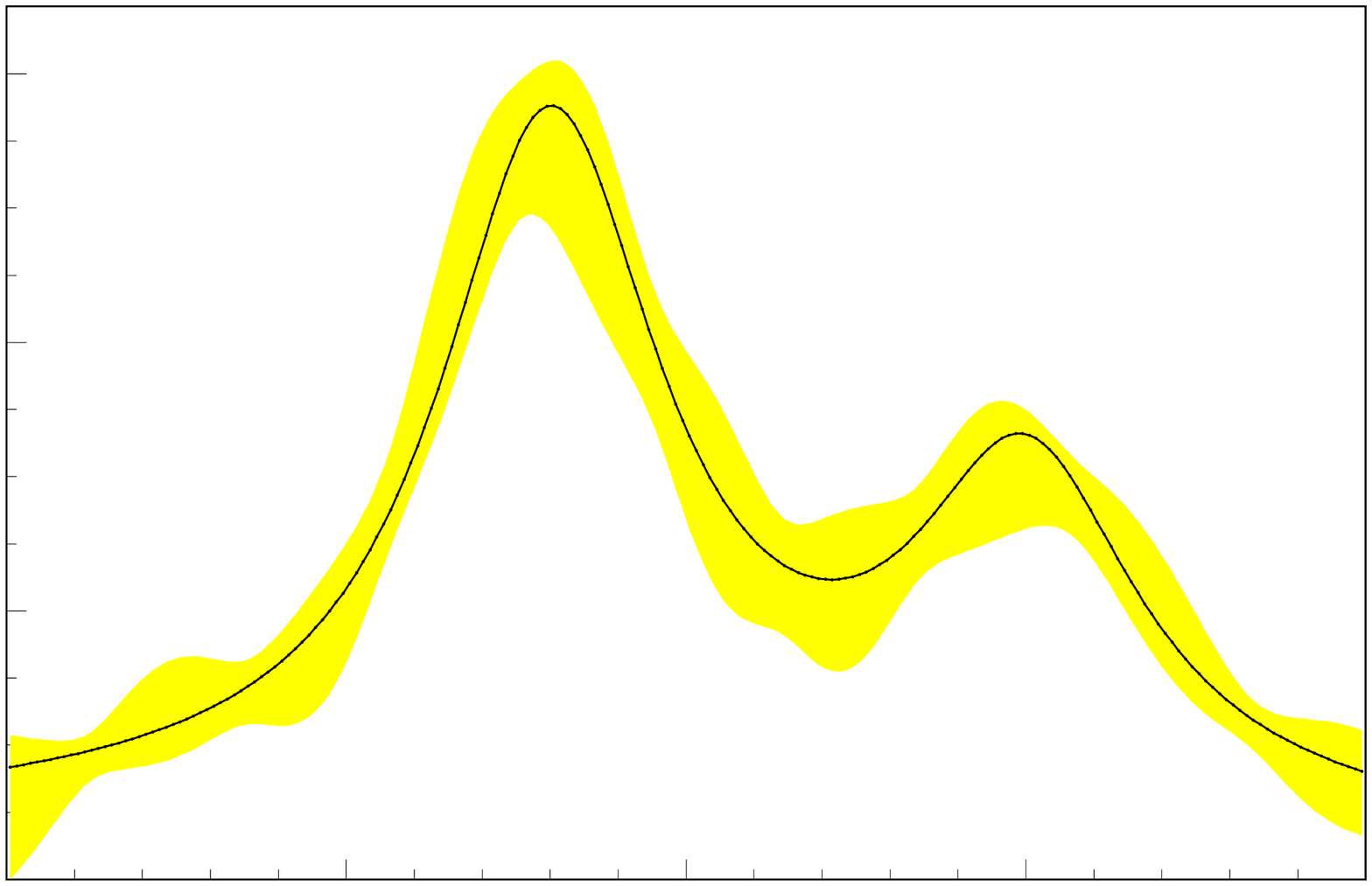}\\
                 \includegraphics[width=2.99in]{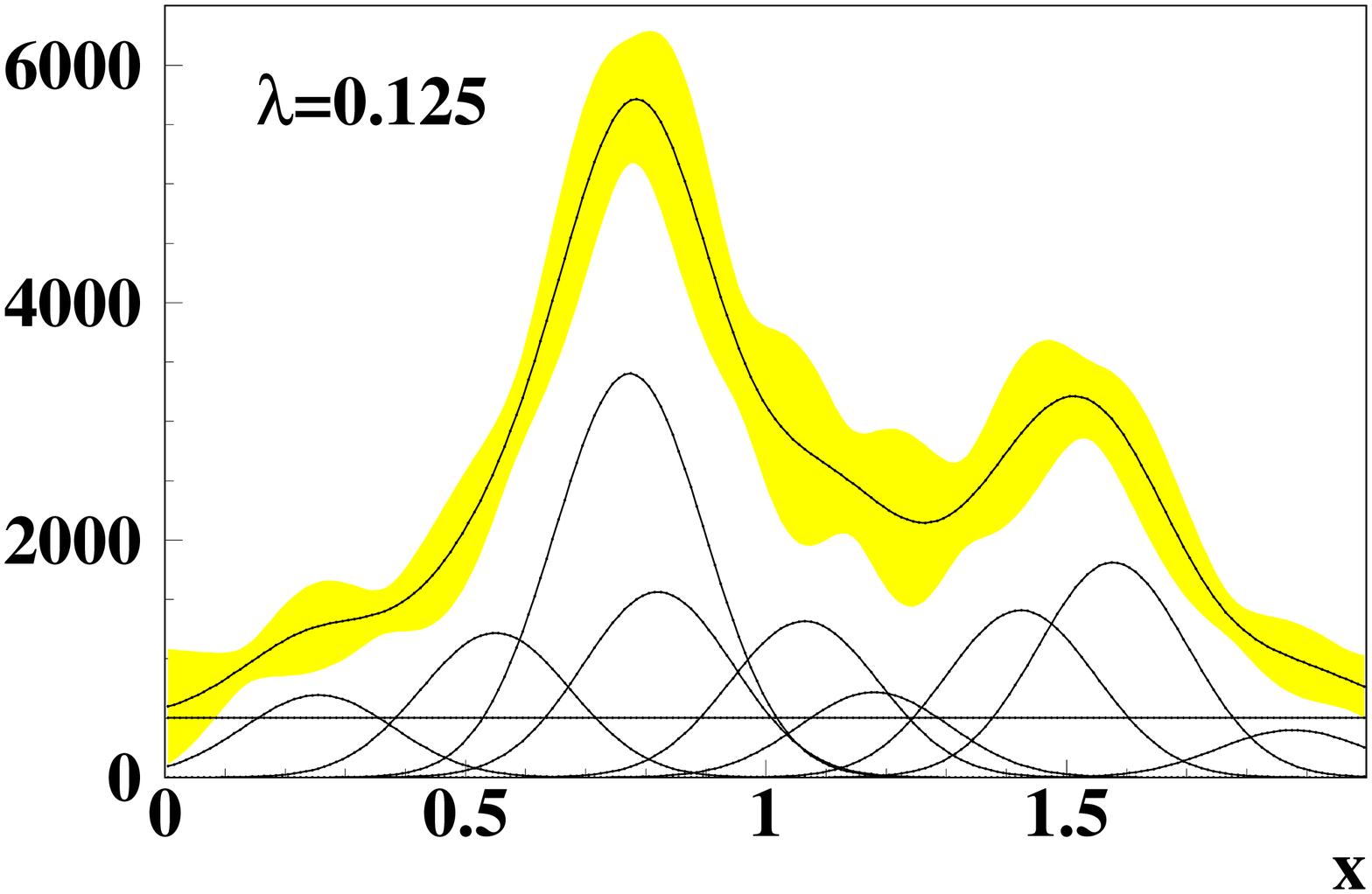} &
\hspace*{-1.82cm}\includegraphics[width=2.99in]{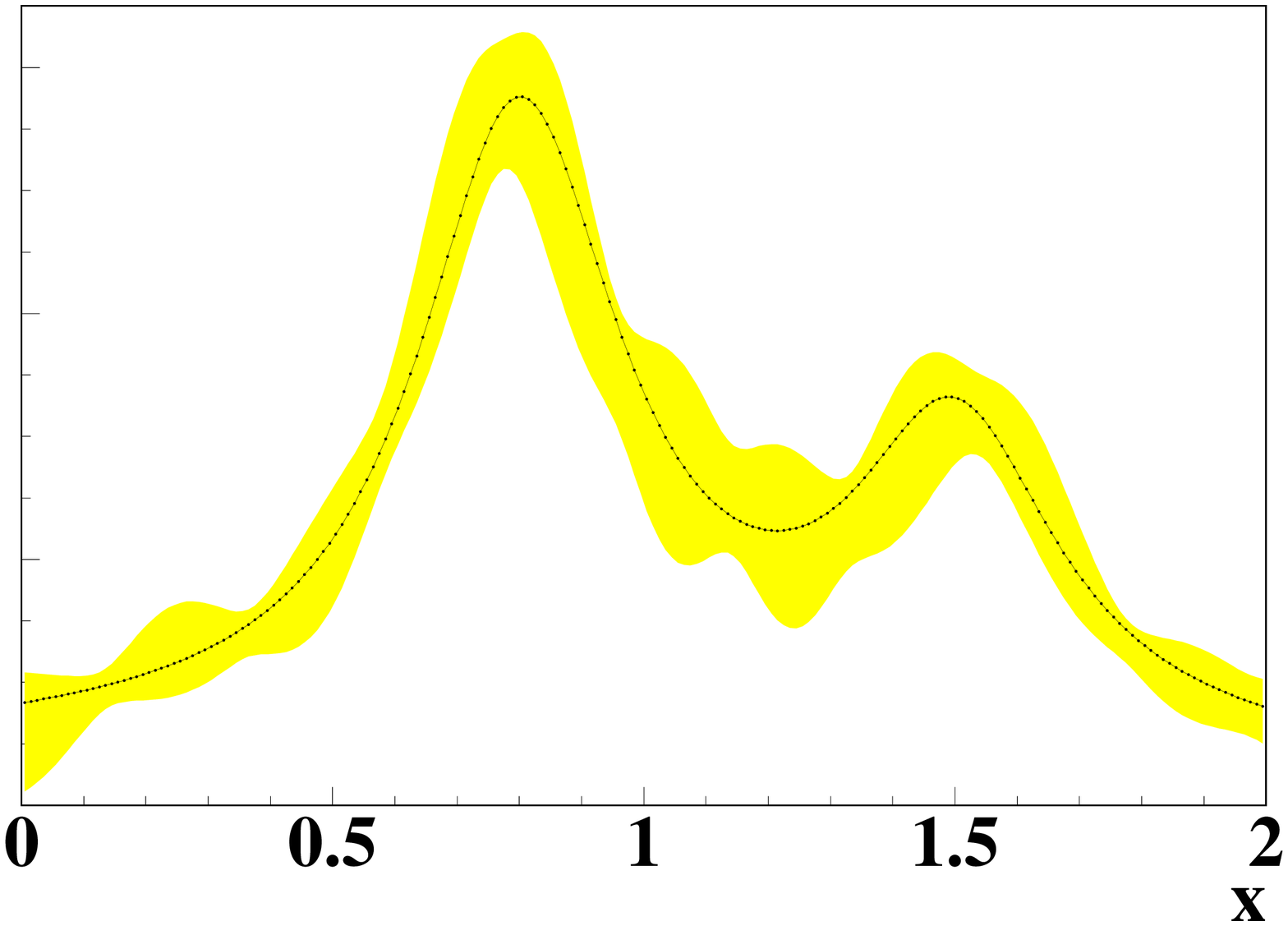}
\end{array}$
\end{center}
\vspace*{-0.3cm}
\caption{Components of the unfolded distribution and the unfolded
         distribution $\hat{p}(x)$\/ given by the sum of the components
         with $\pm 2\delta(x)$\/ interval (left) and the error
         band overlaid with the true distribution $p(x)$\/ (right)
         for different values of the scale parameter $\lambda$.}
\label{fig:numex3}
\end{figure}

\begin{figure}[p]
\centering
\begin{minipage}{0.22\textwidth}
\small
\vspace*{-2cm}
\renewcommand{\arraystretch}{3.7}
\begin{tabular}{ll}
  $\lambda$ & $p$-value \\[-0.8cm]
  0.1   & 0.21  \\
  0.125 & 0.22  \\
  0.15  & 0.20  \\
  0.175 & 0.23  \\
  0.2   & 0.21  \\
  0.225 & 0.05  \\
  0.25  & 0.01  \\
  0.275 & 0.00
\end{tabular}
\end{minipage}
\begin{minipage}{0.75\textwidth}
\includegraphics[width=0.34\textwidth,height=15.0cm]{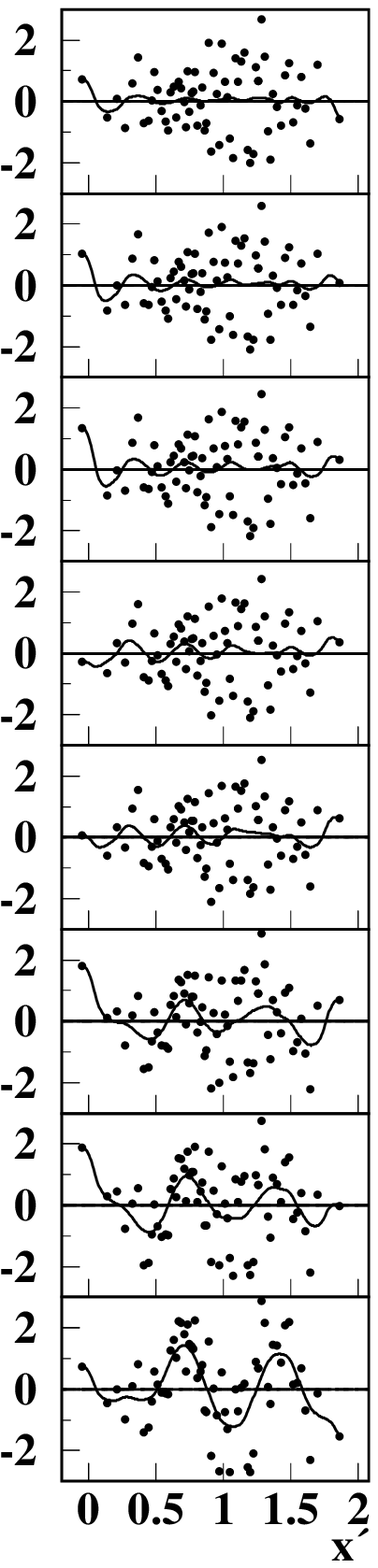}
\raisebox{2.5mm}{\includegraphics[width=0.26\textwidth,height=14.75cm]{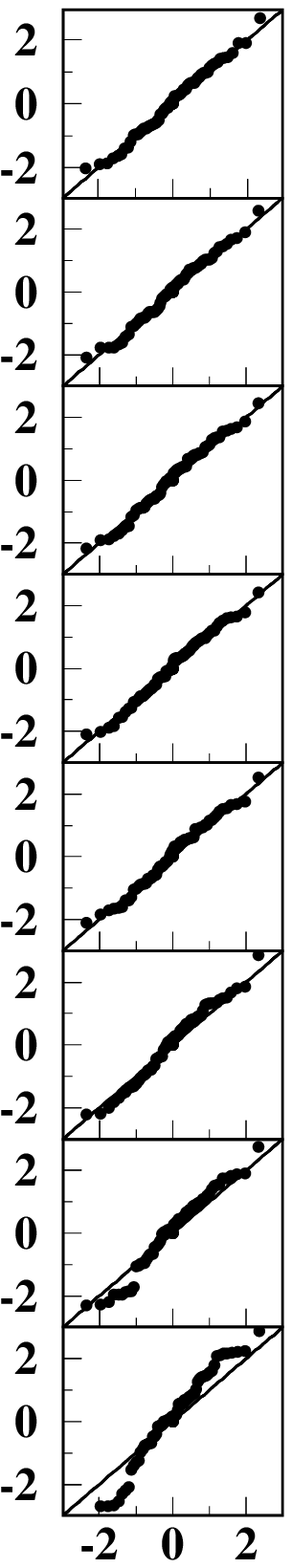}}
\includegraphics[width=0.36\textwidth,height=15.0cm]{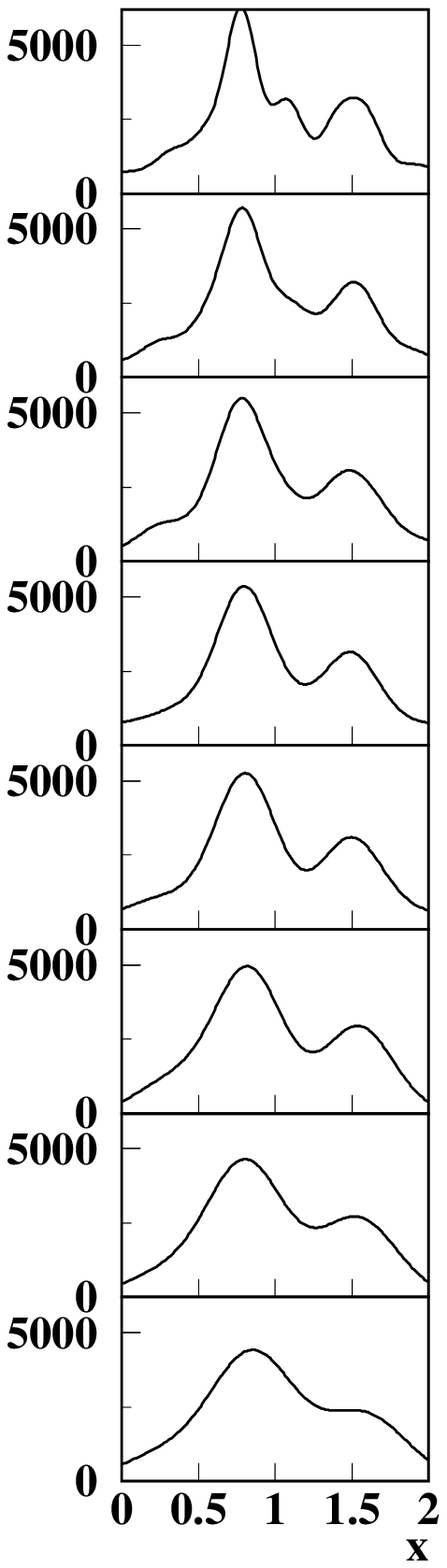}
\end{minipage}
\caption{The $p$-values, residuals, Q-Q plots and unfolded distributions
         for different values of the scale parameter $\lambda$.}
\label{fig:numex4}
\end{figure}

Table\,\ref{tab:result1} gives the results for a scale parameter
$\lambda=0.175$\/ of the gaussian kernels. To illustrate the effect of this
parameter, Fig.\,\ref{fig:numex3} shows how the unfolding results varies with
$\lambda$. The components of the unfolding results are shown together with the
estimate $\hat{p}(x)$. Also shown are the error bands $\pm
2\sqrt{\mathrm{var}[\hat{p}(x)]}$\/ compared to the true distribution $p(x)$.
Figure\,\ref{fig:numex4} illustrates for an even larger range of $\lambda$\/ how
the fit quality varies with the scale paramater. One clearly sees that large
values $\lambda>0.2$\/ lead to a bad fit with a $p$-value $p<0.05$. Here also
significant structures in the residuals and in the quantile-quantile plots are
observed. The smallest value shows some indication of overfitting. The best
parameters for this example evidently are in the range between $0.15<\lambda<0.20$.

This is confirmed when doing a most simple leave-one-out cross validation,
removing in turn each bin of the measured distribution and calculating the
predicted residual sum of squares \cite{allen} as a function of $\lambda$
\begin{equation}
   X_{pr}^2= \sum^n_{i=1} \frac{(P_i- \hat P_{(i)})^2}{\sigma_i^2} \;.
\end{equation}
Here $\hat P_{(i)}$\/ is the estimator for the content of the $i$th bin of the
observed distribution $\bm{P}$, calculated by excluding this bin from the
unfolding procedure or from the determination of the weights for the kernels
selected by the unfolding procedure. The results of the calculation of
$X_{pr}^2/n$\/ for different scale parameters $\lambda$\/ is given in
Tab.\,\ref{tab:xval}. The minimal value of  $X_{pr}^2/n=1.29$\/ is achieved
for $\lambda=0.2$. The choice of $\lambda=0.175$\/ with  $X_{pr}^2/n=1.32$\/
gives a solution with $\sim 20\,\%$\/ larger statistical errors than for
$\lambda=0.2$\/ but, as will be discussed in more detail below, has a
lower bias. The solutions with  $\lambda<0.15$\/ can be considered as
overfitting the data while $\lambda\geq 0.20$ underfits them.

\begin{table}[htb]
\centering
\caption{The $p$-values and average predicted residual sum of squares
         $X_{pr}^2/n$\/ for different values of the scale parameter $\lambda$.}
\label{tab:xval}
\begin{tabular}{c|c|c|c|c|c|c|c|c}
 $\lambda$   & 0.1  & 0.125 & 0.15 & 0.175 & 0.2  & 0.225 & 0.25 & 0.275 \\
\hline
 $p$-value   & 0.21 & 0.22  & 0.20 & 0.23  & 0.21 & 0.05  & 0.01 & 0.00  \\
\hline
 $X_{pr}^2/n$ & 1.77 & 1.42  & 1.46 & 1.32  & 1.29 & 1.58  & 1.84 & 2.02  \\
\end{tabular}
\end{table}

To investigate the statistical properties of the unfolding procedure,
$M=1000$\/ simulation runs were performed producing statistically independent
measured histograms, each based on $N=5000$\/ events for the same true
distribution (\ref{testform}). The unfolded distribution was calculated for
each measured distribution. For the comparison between the unfolding results
and the true distribution a histogram representation is used with $m=12$\/
and alternatively $m=40$ bins. Bin contents are normalized to the bin
width in order to make the bin contents independent of the binning.
The following quantities are considered for each bin $i$\/ of the
unfolded distribution.

\begin{itemize}
\item $p_i$: exact value of bin $i$\/ of the true distribution
  \begin{equation*}
    p_i = \frac{N}{x_i-x_{i-1}} \int_{x_{i-1}}^{x_i} p(x)\,dx
  \end{equation*}
\item $\bar{\hat{p}}_i$: run-averaged value of bin $i$\/ of the unfolded distribution
  \begin{equation*}
    \bar{\hat{p}}_i = \frac{1}{M\cdot(x_{i}-x_{i-1})}\,\sum_{j=1}^{M}\,\hat{p}_i(j)
  \end{equation*}
\item $\mathrm{B}[\hat{p}_i]$: bias in bin $i$\/ of the unfolded distribution
  \begin{equation*}
    \mathrm{B}[\hat{p}_i]= \bar{\hat p}_i - p_i
  \end{equation*}
\item $s_i$: run-averaged standard deviation for bin $i$\/
  \begin{equation*}
    s^2_i= \frac{1}{M-1}\sum_{j=1}^{M}(\hat{p}_i(j)-\bar{\hat{p}}_i)^2
  \end{equation*}
\item $\delta_i$: run-averaged error estimate for bin $i$\/
  \begin{equation*}
    \bar{{\delta}}_{i}= \frac{1}{M\cdot(x_{i}-x_{i-1})}\,\sum_{j=1}^{M} \delta_{i}(j)
  \end{equation*}
\item $\mathrm{B}[\delta_{i}]$: bias on the error of bin $i$\/
  \begin{equation*}
    \mathrm{B}[\delta_{i}] = \bar{\delta}_{i}- s_i
  \end{equation*}
\item $\mathrm{RMSE}_i$: run-averaged Root Mean Square Error for bin $i$
  \begin{equation*}
    \mathrm{RMSE}^2_i
    = \frac{1}{M} \sum_{j=1}^{M} (\hat{p}_i(j)-p_i)^2
    =  s_i^2 +  \mathrm{B}[p_i]^2
  \end{equation*}
\end{itemize}

In addition to the bin-dependent quantities some global measures
for the quality of the unfolding result are defined by
summing over all $m$\/ bins of the unfolded distribution.

\begin{itemize}
\item $\mathrm{TRMSB}$: Total Root Mean Square Bias
  \begin{equation*}
    \mathrm{TRMSB} = \sqrt{\frac{1}{m}\sum_{i=1}^m \mathrm{B}[p_i]^2}
  \end{equation*}
\item $\mathrm{TRMSV}$: Total Root Mean Square Variance
  \begin{equation*}
    \mathrm{TRMSV} = \sqrt{\frac{1}{m}\sum_{i=1}^m s_i^2}
  \end{equation*}
\item $\mathrm{TRMSE}$: Total Root Mean Square Error
  \begin{equation*}
    \mathrm{TRMSE} = \sqrt{\frac{1}{m} \sum_{i=1}^m \mathrm{RMSE}_i^2}
                   = \sqrt{\mathrm{TRMSB}^2+\mathrm{TRMSV}^2}
  \end{equation*}
\end{itemize}

\begin{figure}[p]
\vspace*{-2.cm}
\begin{center}$
\begin{array}{cc}
\vspace *{-1.65cm}\includegraphics[width=2.5in]{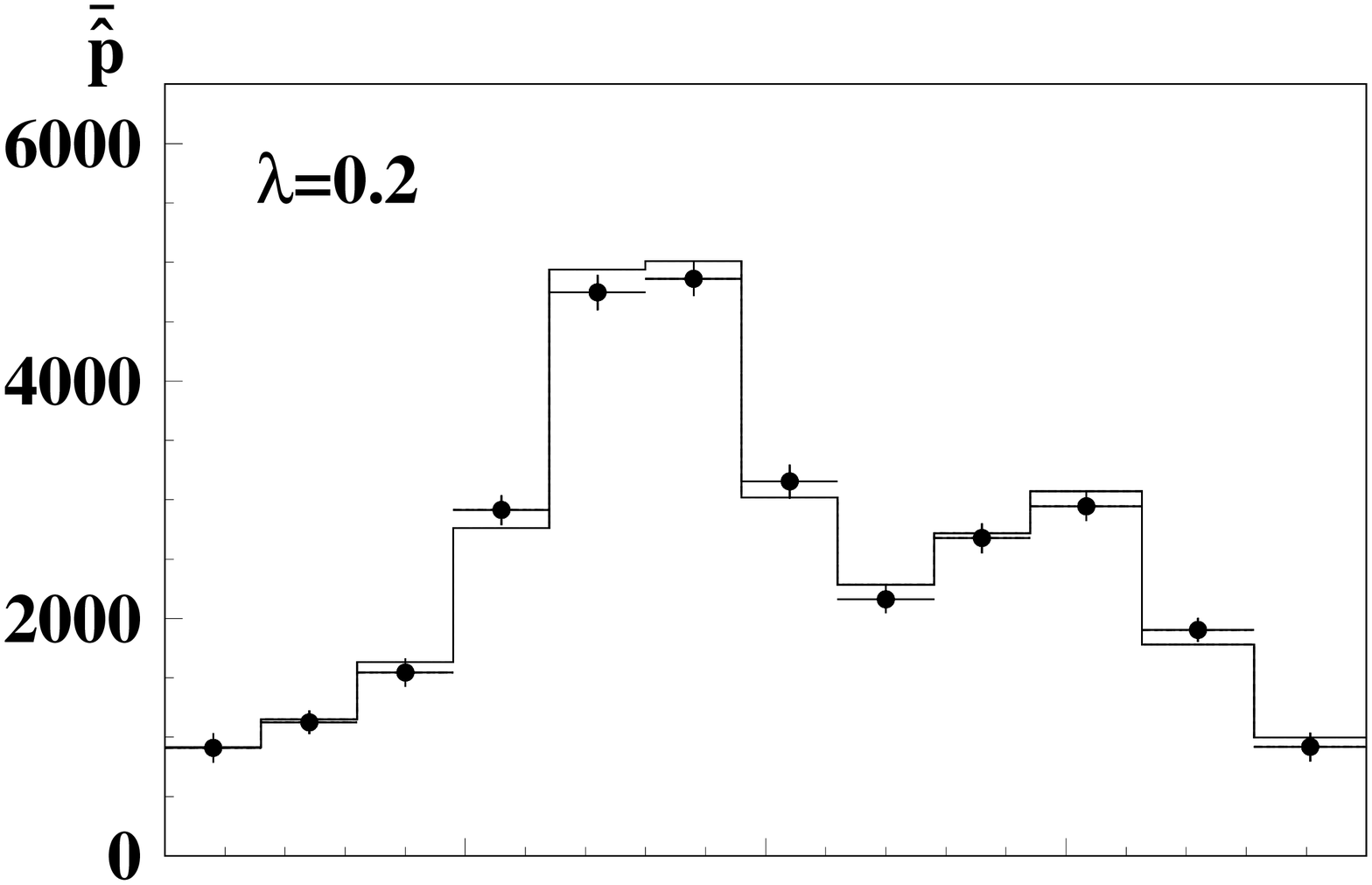} &
 \hspace *{-1.6cm}\includegraphics[width=2.5in]{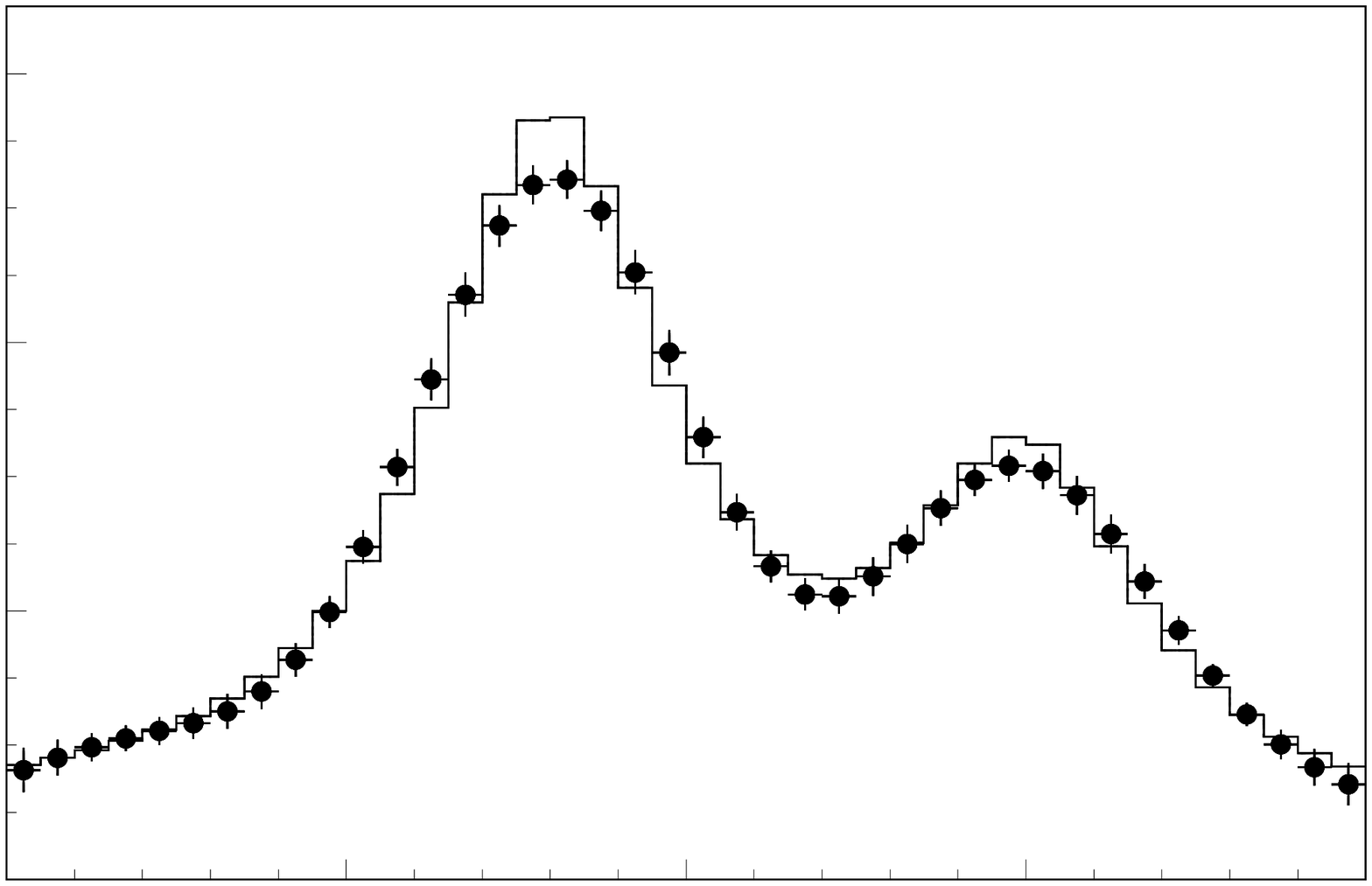}\\
\vspace*{-1.1cm}\includegraphics[width=2.5in]{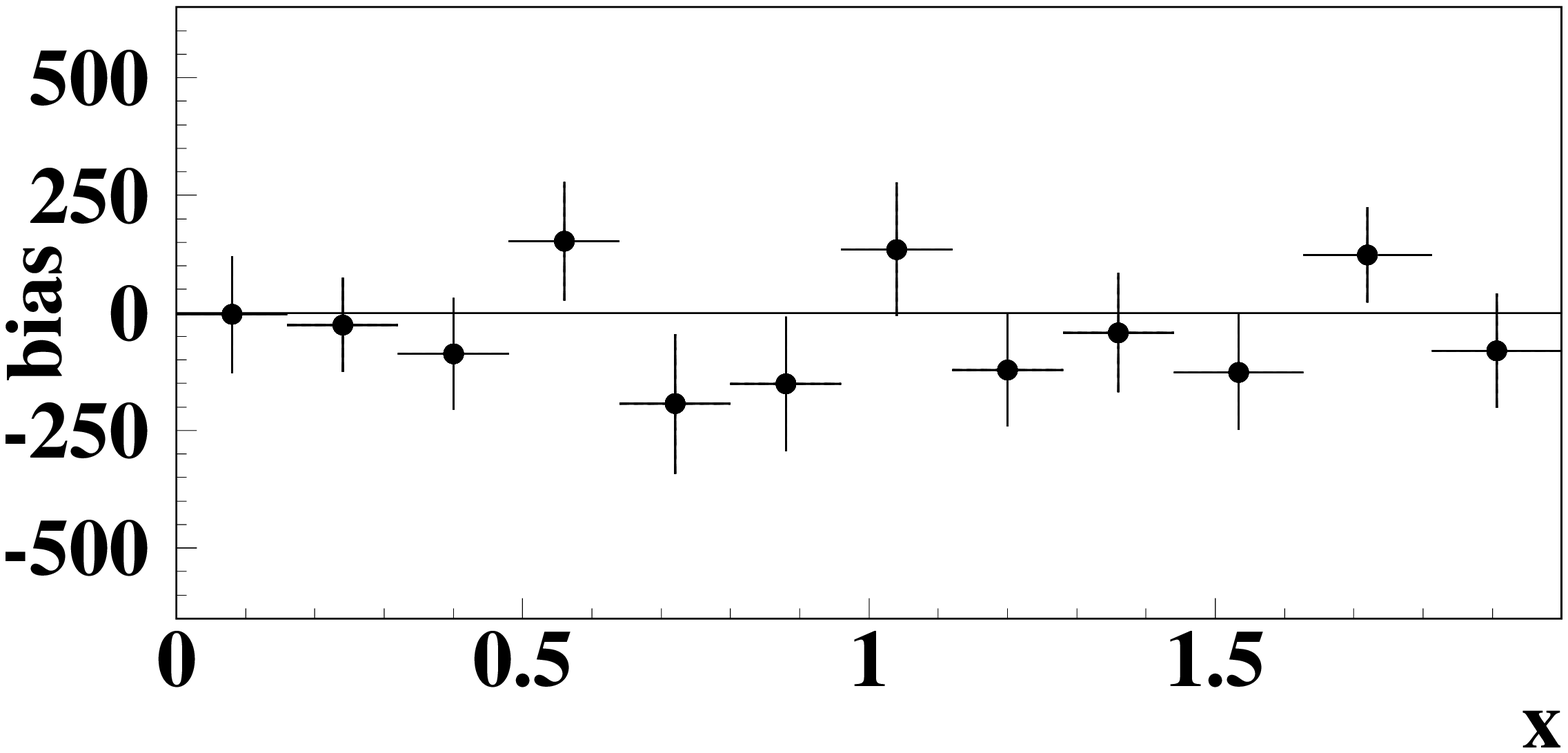} &
 \hspace *{-1.6cm} \includegraphics[width=2.5in]{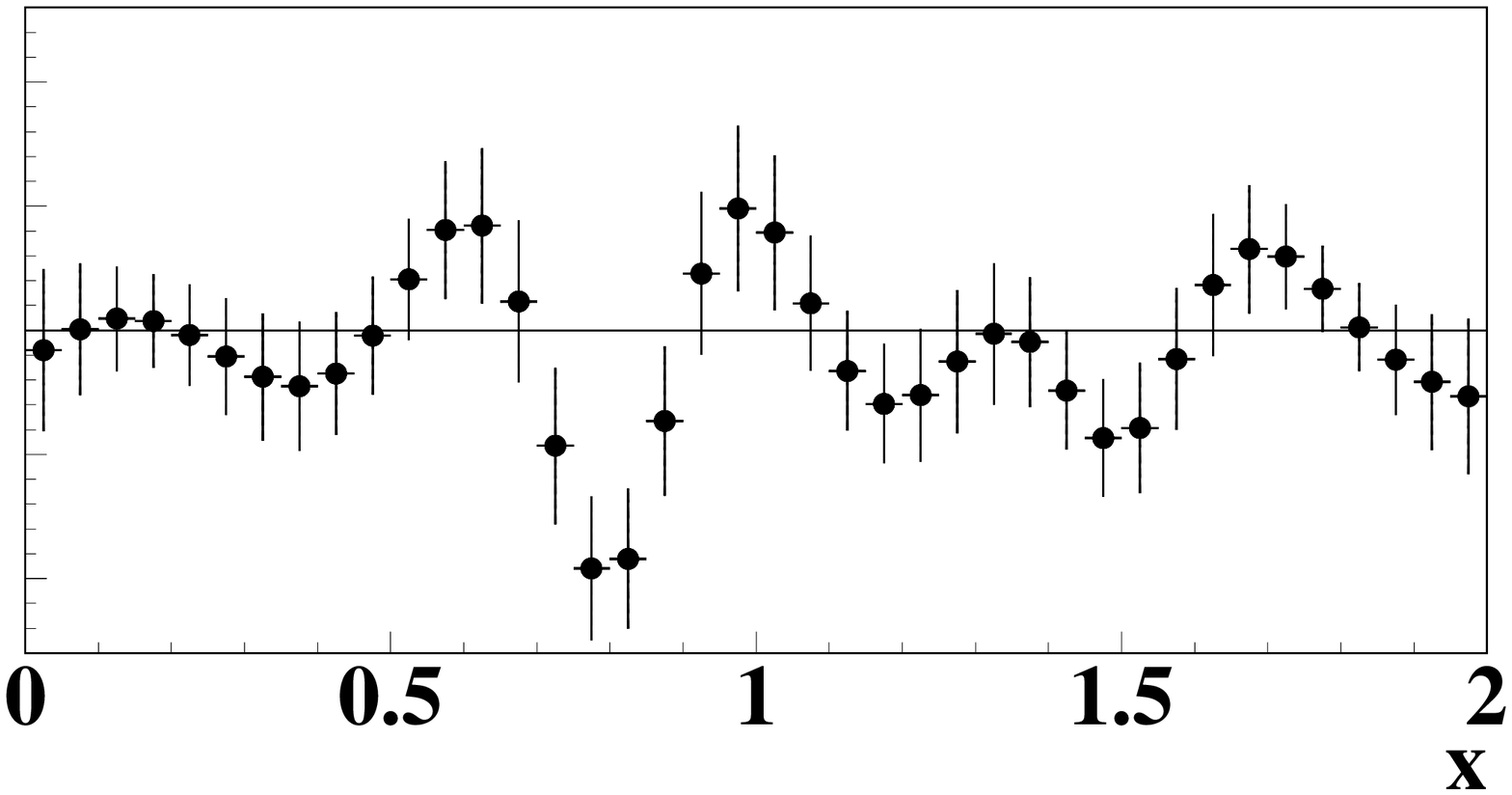}\\
\vspace*{-1.65cm}\includegraphics[width=2.5in]{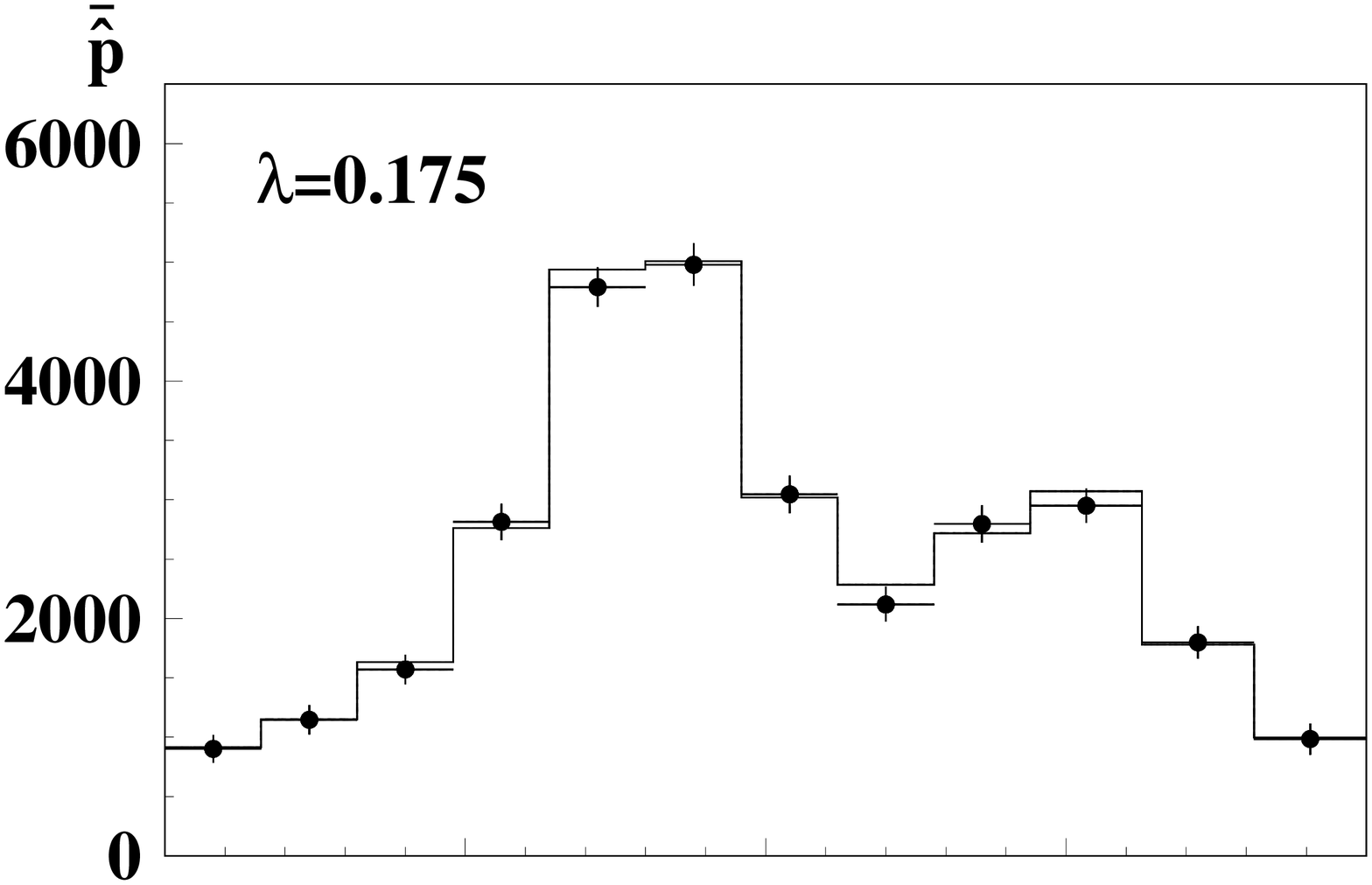} &
 \hspace *{-1.6cm} \includegraphics[width=2.5in]{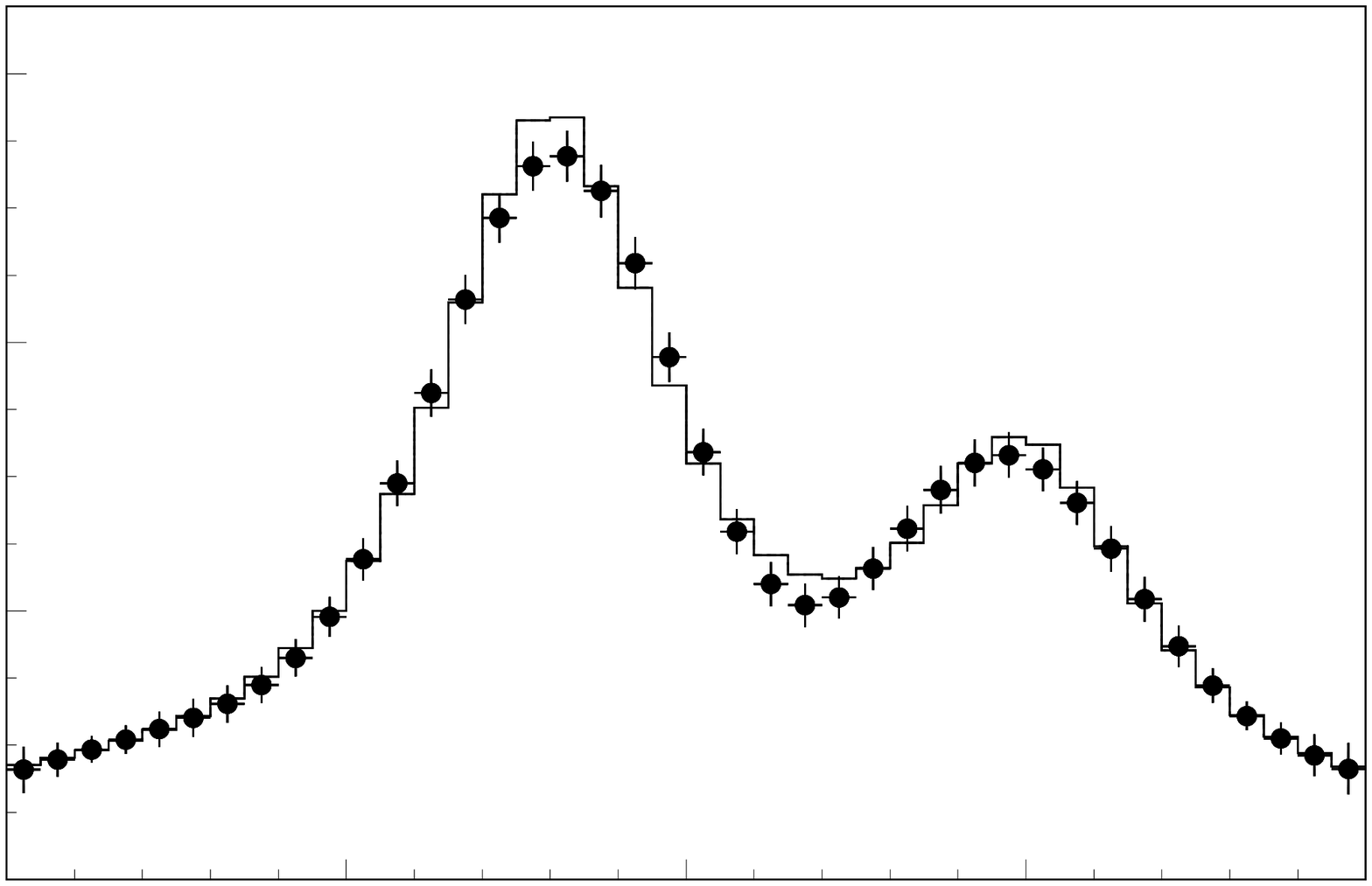}\\
\vspace*{-1.1cm}\includegraphics[width=2.5in]{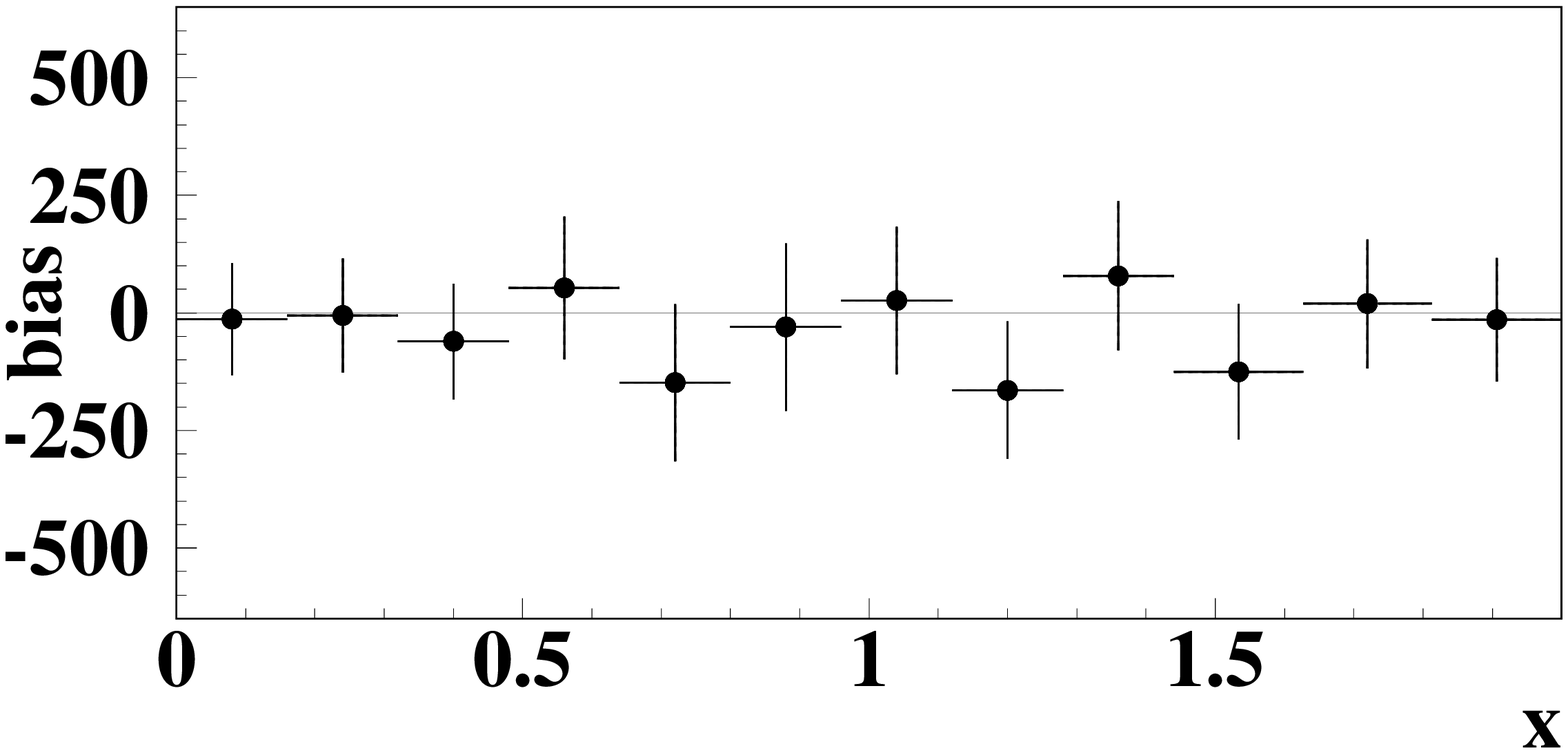} &
 \hspace *{-1.6cm} \includegraphics[width=2.5in]{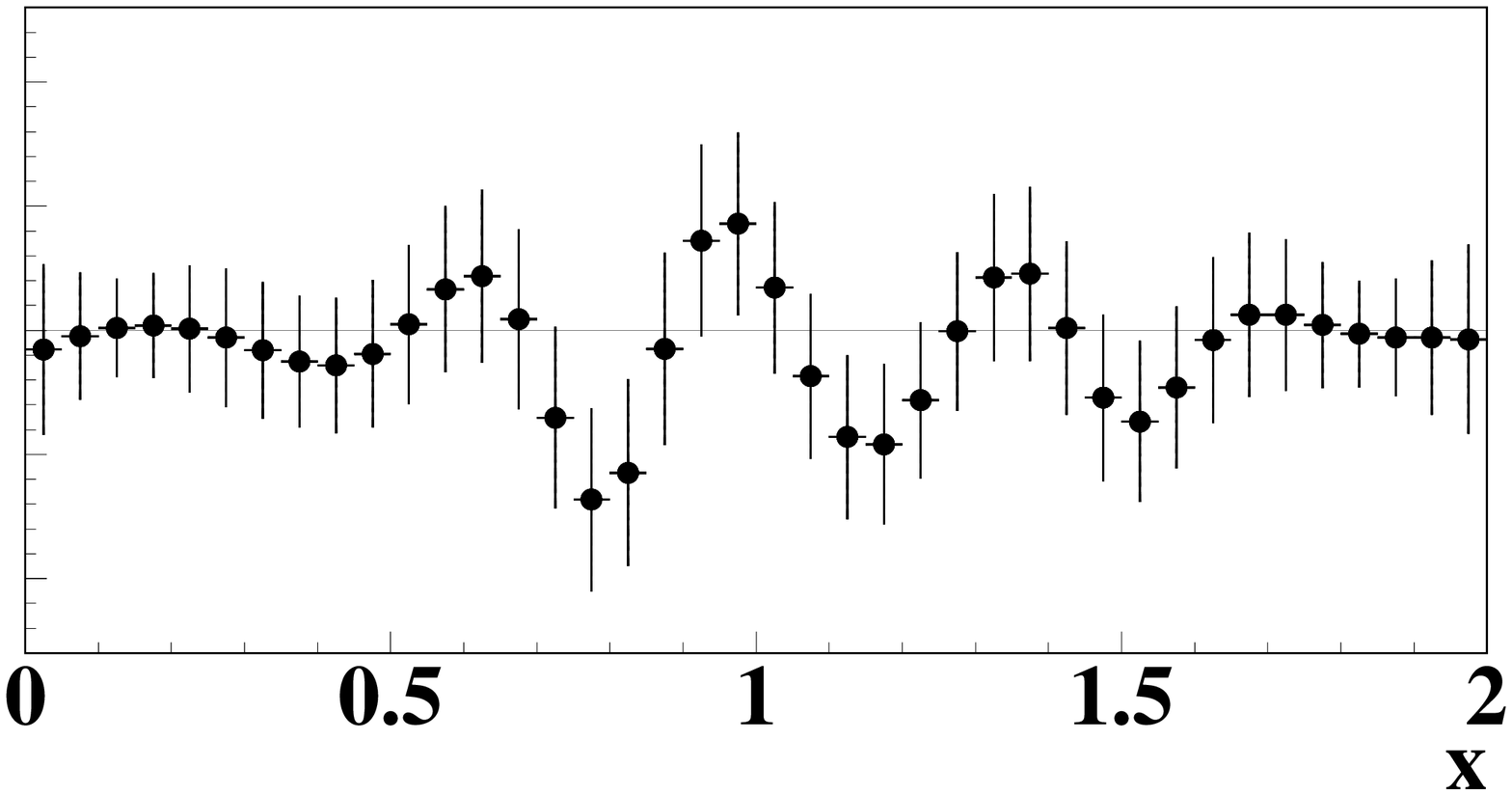}\\
\vspace*{-1.65cm}\includegraphics[width=2.5in]{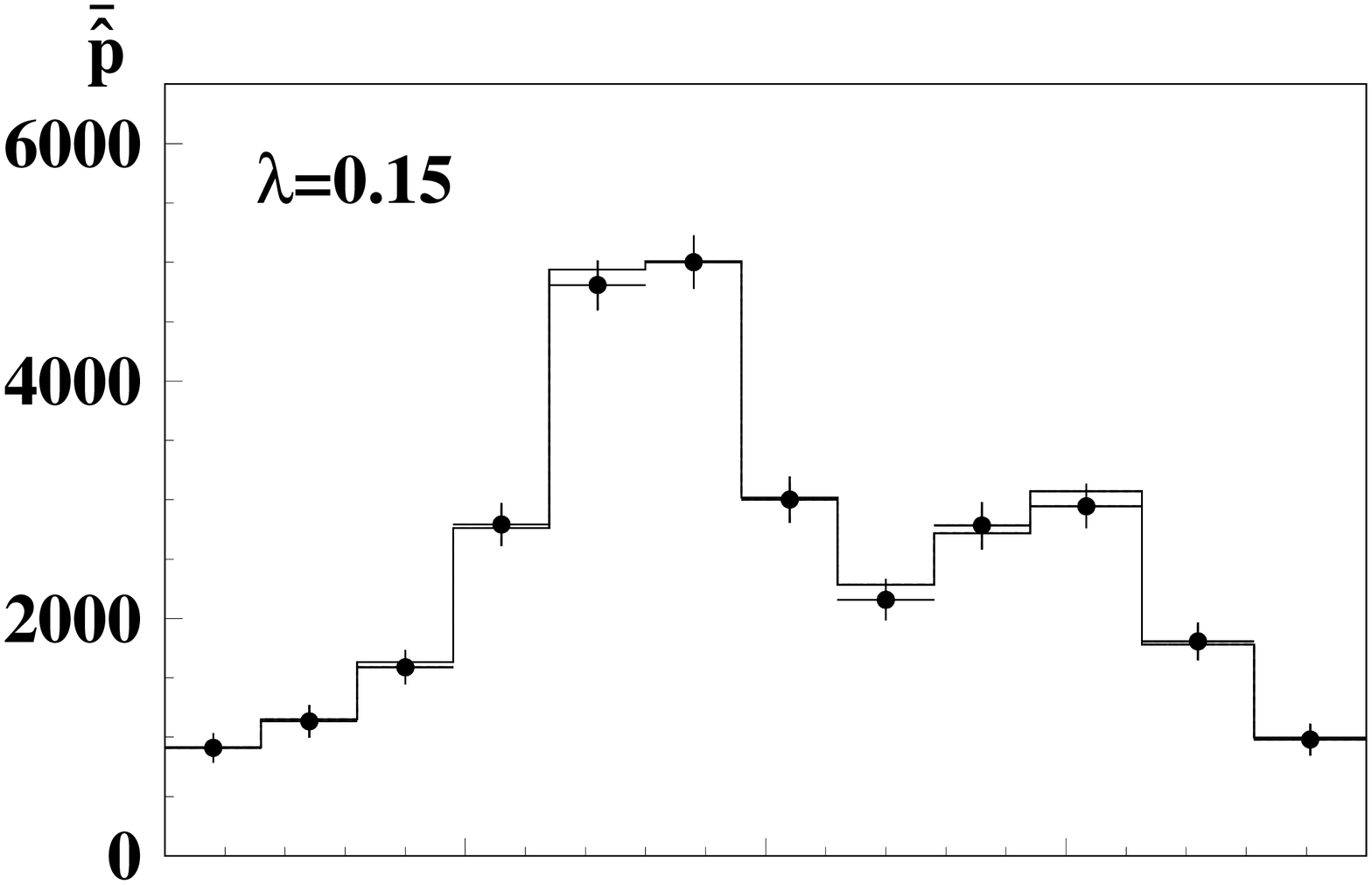} &
 \hspace *{-1.6cm} \includegraphics[width=2.5in]{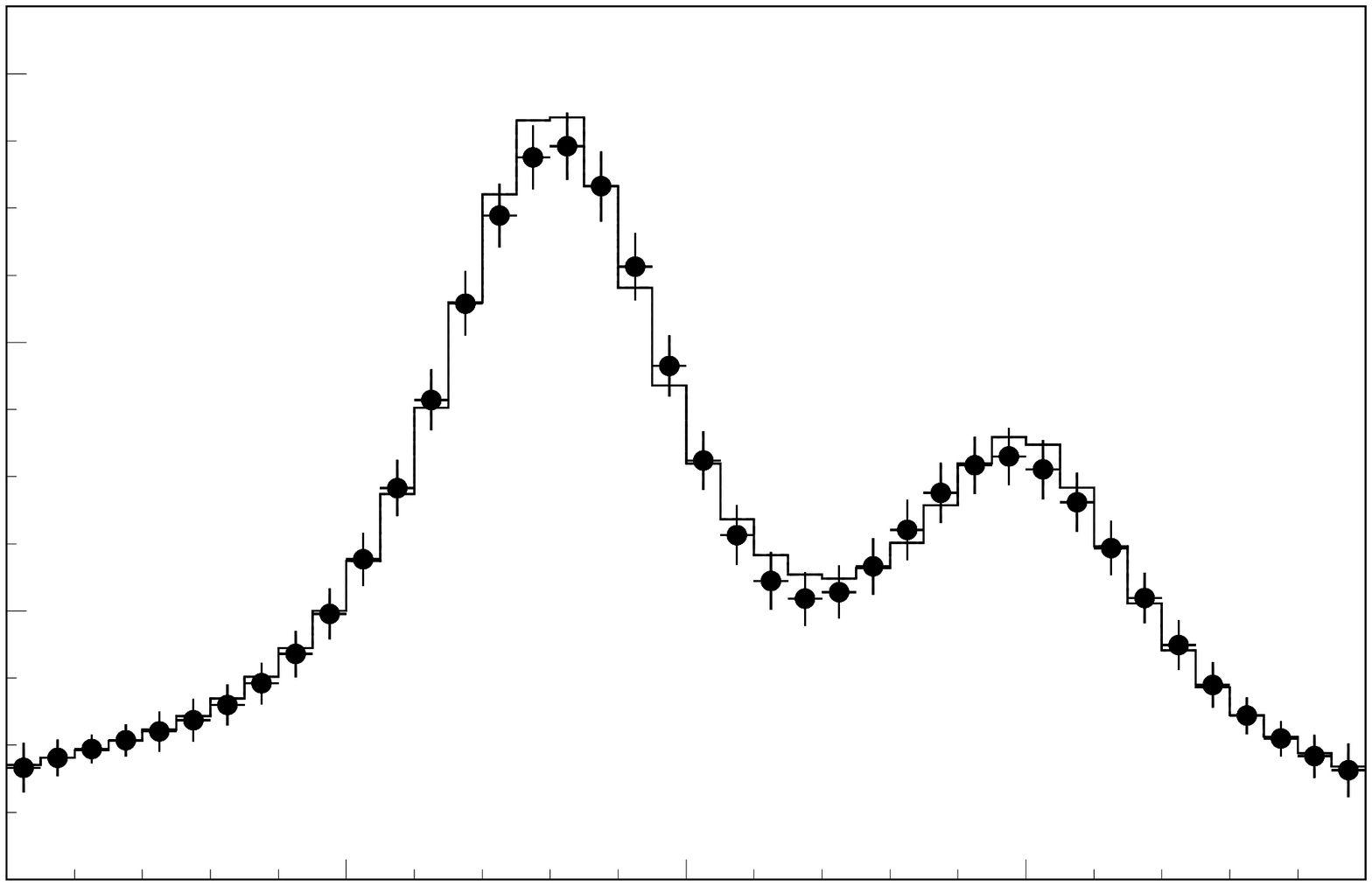}\\
                   \includegraphics[width=2.5in]{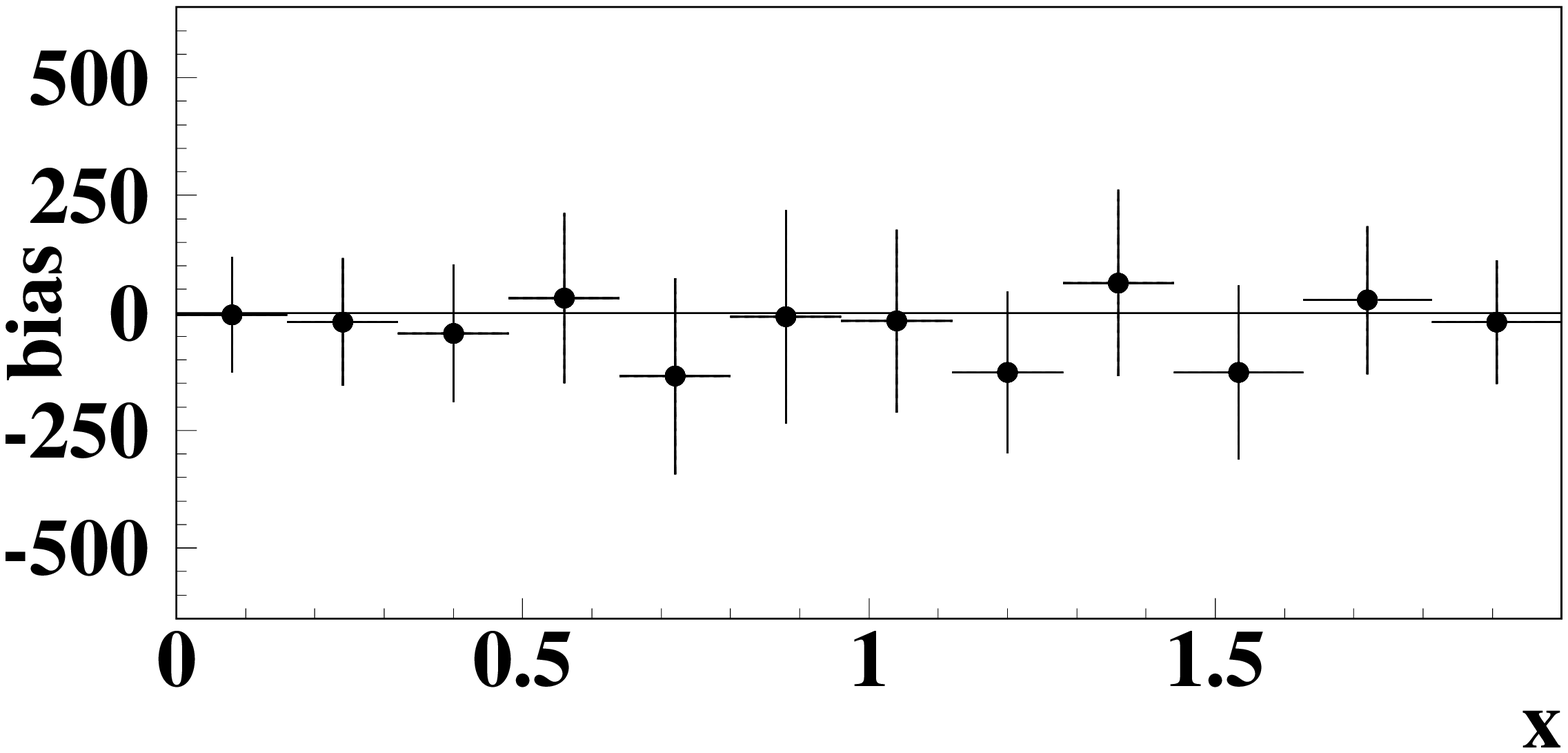} &
 \hspace *{-1.6cm} \includegraphics[width=2.5in]{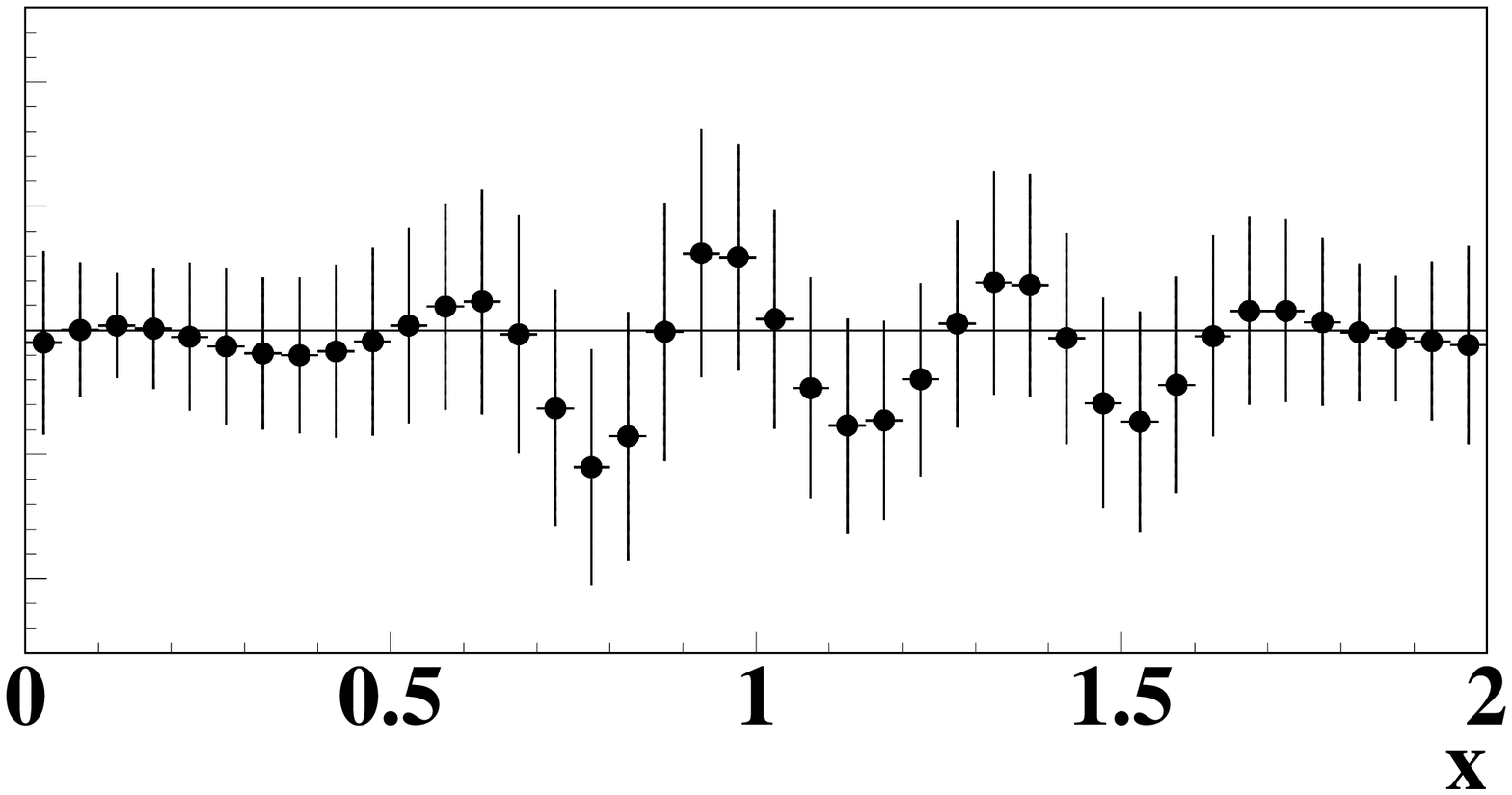}
\end{array}$
\end{center}
\vspace*{-0.3cm}
\caption{Average unfolding results $\bm{\bar{\hat{p}}}$\/ and bias
         $\mathrm{B}[\hat{p}_{i}]$\/ using $\lambda=0.2$, $\lambda=0.175$,
         and $\lambda=0.15$\/  for $m=12$\/ (left) and $m=40$\/ (right) bins.
         The vertical error bars denote the standard deviations  $s_{i}$.
         The histograms show the true bin contents $p_i$.}
\label{fig:avgunf}
\end{figure}

Numerical calculations of the characteristics of the unfolding procedure
for 12 bins and gaussian kernels with $\lambda=0.175$\/ are presented
in Tab.\,\ref{tab:foms}. One sees that the bias is small compared to the
statistical errors of the unfolding result and that the error estimates
agree well with the actual scatter of the results. A visual representation
of these findings for different values $\lambda$\/ is given in
Fig.\,\ref{fig:avgunf} for $m=12$\/ and $m=40$\/ bins of the unfolded
distribution. At the resolution of $12$\/ bins the unfolding result is
consistent with the true distribution, at $40$\/ bins and $\lambda=0.2$\/
one observes some systematic effects in the bias distributions. The bias
gets smaller with decreasing values $\lambda$, but the errors become larger,
which illustrates the well known ''bias-noise complementary law''
\cite{zhigunov2} that the noise grows when the regularization parameter
tends to zero.
\begin{table}[htb]
\centering
\caption{Exact values of the bins of the true distribution $p_i$,
         average values $\bar{\hat{p}}_i$\/ from the unfolding procedure,
         bias $\mathrm{B}[\hat{p}_{i}]$, standard deviation $s_i$,
         mean error $\bar{\delta}_{i}$, bias of the calculated errors
         $\mathrm{B}[\delta_{i}]$\/ and Root Mean Square Errors $RMSE_i$\/
         for $\lambda=0.0175$.}
\label{tab:foms}
\vspace*{2mm}
\begin{tabular}{r|rrrrrrr}
 $i$                     &
 $p_i$                   &
 $\bar{\hat{p}}_i$       &
 $\mathrm{B}[\hat{p}_{i}]$&
 $s_i $                   &
 $\bar{\delta}_{i}$       &
 $\mathrm{B}[\delta_{i}]$ &
 $\mathrm{RMSE}_i$        \\
\hline
  1 &   913. &     900. &     -13. &     119. &     123. &       4. &     120. \\
  2 &  1152. &    1146. &      -5. &     121. &     123. &       2. &     121. \\
  3 &  1631. &    1570. &     -61. &     123. &     129. &       6. &     137. \\
  4 &  2760. &    2813. &      53. &     152. &     156. &       5. &     161. \\
  5 &  4941. &    4793. &    -149. &     167. &     184. &      17. &     223. \\
  6 &  5011. &    4981. &     -30. &     177. &     190. &      13. &     180. \\
  7 &  3018. &    3044. &      26. &     157. &     164. &       7. &     159. \\
  8 &  2284. &    2119. &    -165. &     146. &     157. &      11. &     220. \\
  9 &  2718. &    2797. &      79. &     159. &     167. &       8. &     177. \\
 10 &  3073. &    2948. &    -125. &     144. &     165. &      21. &     191. \\
 11 &  1779. &    1798. &      19. &     136. &     150. &      14. &     138. \\
 12 &   997. &     983. &     -14. &     131. &     127. &      -4. &     132. \\
\end{tabular}
\end{table}

\begin{figure}[htb]
\centering
\includegraphics[width=0.475\textwidth]{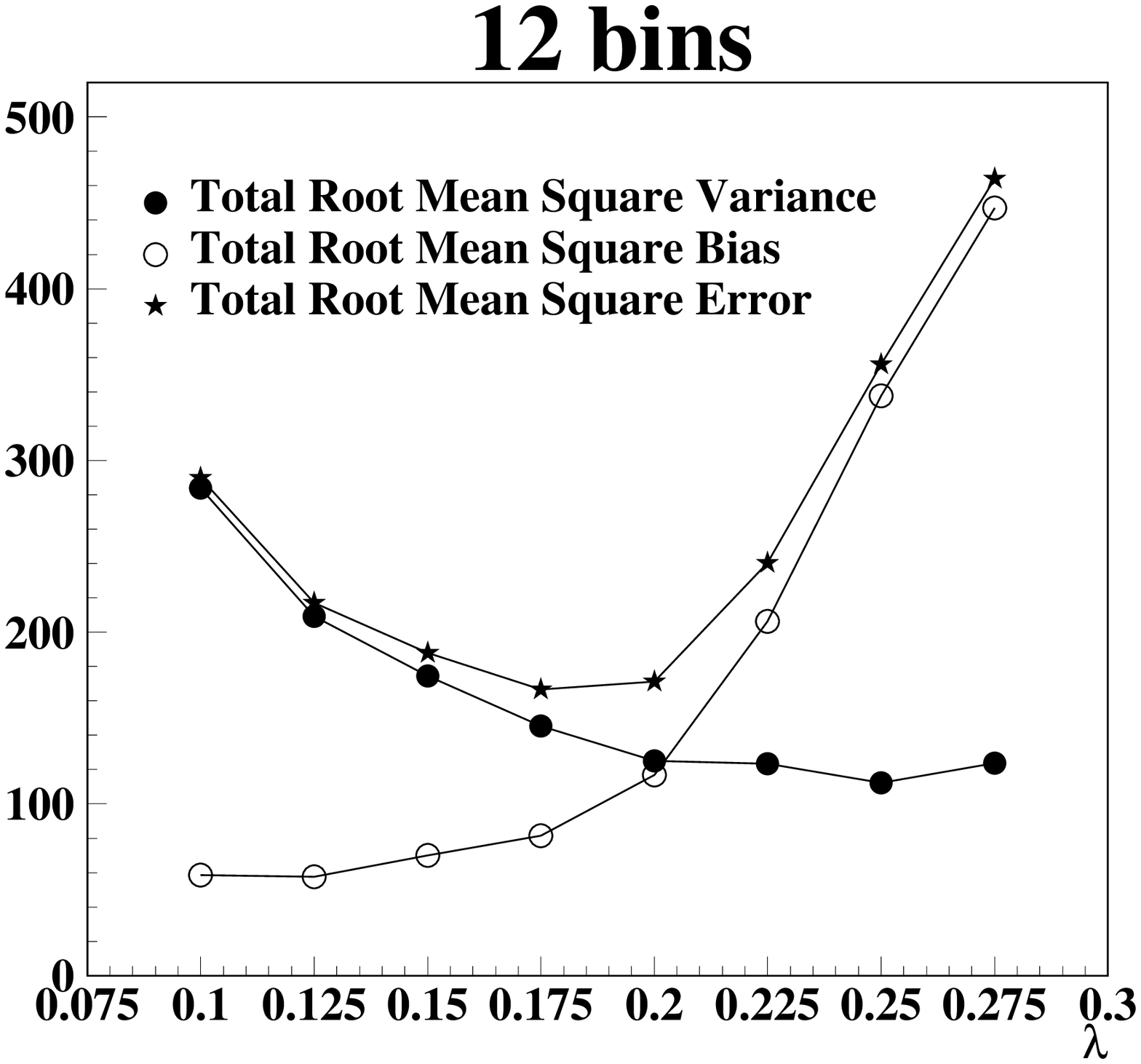}
\includegraphics[width=0.475\textwidth]{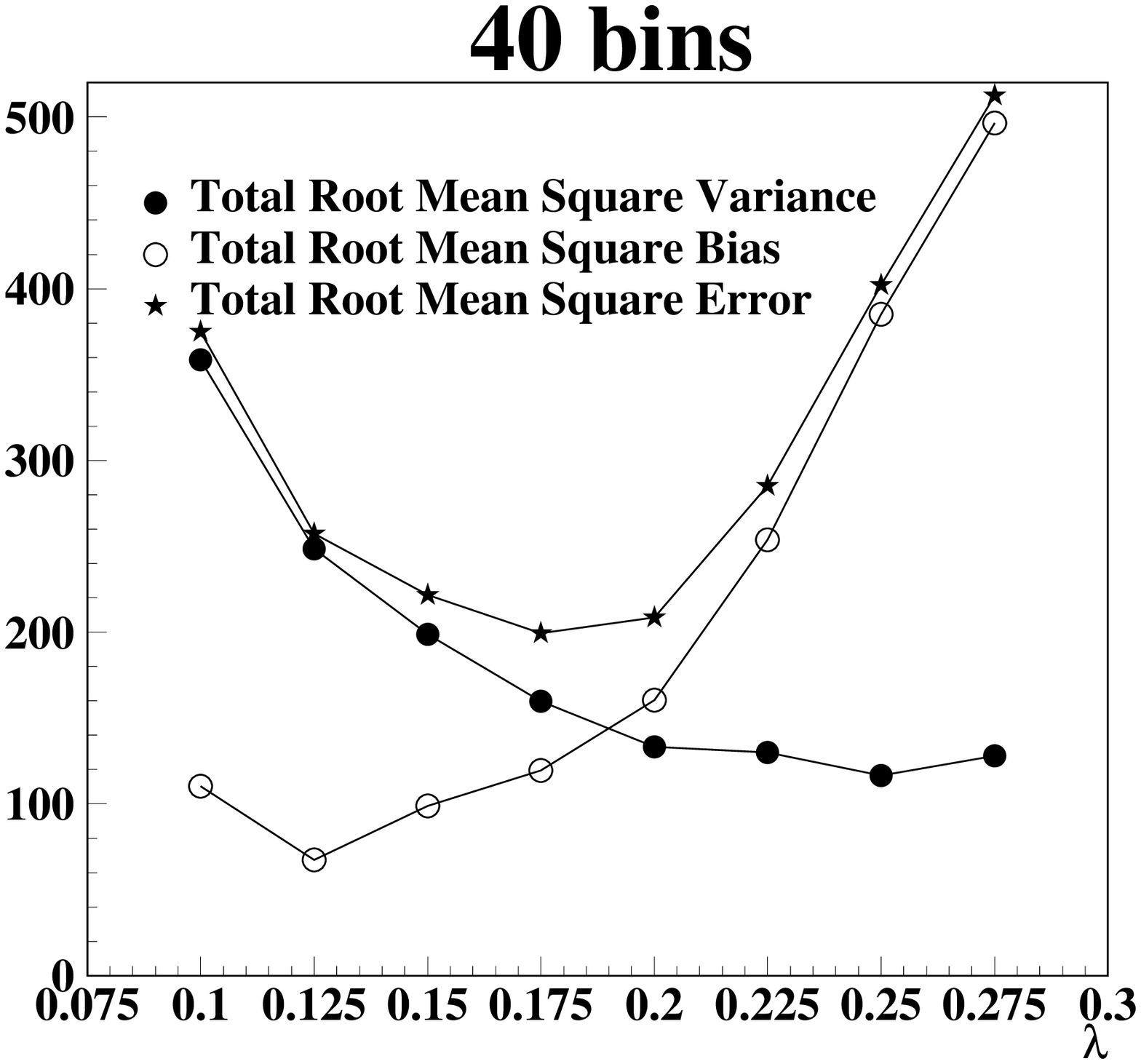}
\caption{Global characteristics of the unfolding result for 12 bins
         (left) and 40 bins (right) as a function of the scale parameter
         $\lambda$.}
\label{fig:globfom}
\end{figure}

The behavior of the global characteristics using 12 or 40 bins for
the unfolded distribution is shown in Fig.\,\ref{fig:globfom}. The
behavior in both cases is very similar. The plots show how with
increasing scale parameter $\lambda$, i.e. stronger regularization,
statistical errors decrease while the bias increases. Adding both
contributions in quadature, the Total Root Mean Square Error shows
a minimum around $\lambda=0.175$, i.e. in the region also favored by
the cross-validation approach for the determination of $\lambda$.
% Minimal value of Total Root Mean Square Error (TRMSE) for the unfolding
% proble%m can not be lower then the TRMSE of  the fitting of  the true
% distribution by %particular kernel for the  given number of events. The  best
% $\lambda$ for unf%olding problem can not be lower then  best $\lambda$ for
% fitting  true distribu%tion by kernels.
\section{Conclusions}
\label{sec:conclusions}
A new method for unfolding the true distribution from experimental data is
presented. The unfolding problem is known as an ill-posed problem which
can not be solved without some a priori information about solution. Smoothness
and positiveness are examples for this type of information. In the
proposed algorithm the unknown true distribution is represented as
a weighted sum of smooth kernels. The scale parameter of the kernels acts as
a regularization parameter allowing to adjust the smoothness of the result.
A cross-validation approach is proposed to determine an optimal value
of this parameter. The method avoids discretization of the integral
equation which is often done by unfolding methods and is an additional
source of bias for the solution of unfolding problem. Various criteria
were discussed to gauge the quality of the unfolding result. The methods
provides a solution for the unfolding problem with a non-singular error
matrix which can be used to test the consistency of a theoretical prediction
with the experimental data. A numerical example including extensive simulation
studies of the statistical properties of the method was presented to illustrate
and to validate the procedure. For the example typical execution times per
unfolding were found to be around 0.1\,s on a 2 GHz CPU. The method can be
extended to deal with steeply falling spectra or multidimensional
distributions and to handle properly the case of limited statistics in the
determination of the response function.
\\

{\bf{Acknowledgements}}
\\

\noindent
The authors are grateful to Markward Britsch for useful
 discussions and careful reading of the manuscript. One of
 us (NG) thanks the University of Akureyri and the MPI for
 Nuclear Physics for support in carrying out the research.

\end{document}